\documentclass[a4paper,fleqn,usenatbib]{mnras}
\usepackage{newtxtext,newtxmath}
\usepackage[T1]{fontenc}
\usepackage{ae,aecompl}
\usepackage{graphicx}
\usepackage{longtable}

\newcommand{\kms}{\mbox{$\mathrm{km\,s^{-1}}$}}
\newcommand{\MSUN}{\,\mbox{$\mathrm{M_{\odot}}$}}
\newcommand{\RSUN}{\,\mbox{$\mathrm{R_{\odot}}$}}
\newcommand{\Ion}[2]{#1{\,\sc#2}}

\title[Testing the white dwarf mass-radius relation]{Testing the white dwarf
  mass-radius relationship with eclipsing binaries}

\author[S. G. Parsons et al.]{S.~G.~Parsons$^{1}$\thanks{s.g.parsons@sheffield.ac.uk},
B.~T.~G{\"a}nsicke$^{2}$,
T.~R.~Marsh$^{2}$,
R.~P.~Ashley$^{2}$,
M.~C.~P.~Bours$^{3}$
\newauthor
E.~Breedt$^{2}$,
M.~R.~Burleigh$^{4}$,
C.~M.~Copperwheat$^{5}$,
V.~S.~Dhillon$^{1,6}$,
M.~Green$^{2}$,
\newauthor
L.~K.~Hardy$^{1}$,
J.~J.~Hermes$^{7}$,
P.~Irawati$^{8}$,
P.~Kerry$^{1}$,
S.~P.~Littlefair$^{1}$,
\newauthor
M.~J.~McAllister$^{1}$,
S.~Rattanasoon$^{1,7}$,
A.~Rebassa-Mansergas$^{9}$,
D.~I.~Sahman$^{1}$
\newauthor
and M.~R.~Schreiber$^{3,10}$
\\
$^{1}$ Department of Physics and Astronomy, University of Sheffield,
Sheffield, S3 7RH, UK\\
$^{2}$ Department of Physics, University of Warwick, Coventry CV4 7AL, UK\\
$^{3}$ Instituto de F{\'i}sica y Astronom{\'i}a, Universidad de
Valpara{\'i}so, Avenida Gran Bretana 1111, Valpara{\'i}so, 2360102, Chile\\
$^{4}$ Department of Physics and Astronomy, University of Leicester, Leicester
LE1 7RH, UK\\
$^{5}$ Astrophysics Research Institute, Liverpool John Moores University, IC2,
Liverpool Science Park, L3 5RF, UK\\  
$^{6}$ Instituto de Astrof{\'i}sica de Canarias, V{\'i}a Lactea s/n, La
Laguna, E-38205 Tenerife, Spain\\
$^{7}$ Hubble Fellow, Department of Physics and Astronomy, University of 
North Carolina, Chapel Hill, NC 27599-3255, USA\\
$^{8}$ National Astronomical Research Institute of Thailand, 191 Siriphanich
Bldg., Huay Kaew Road, Chiang Mai 50200, Thailand\\
$^{9}$ Departament de F\'isica, Universitat Polit\'ecnica de Catalunya,
c/Esteve Terrades 5, 08860 Castelldefels, Spain\\
$^{10}$ Millenium Nucleus "Protoplanetary Disks in ALMA Early Science",
Universidad de Valparaiso, Valparaiso 2360102, Chile\\
}

\date{Accepted 2017 June 15. Received 2017 June 15; in original form 2017 February 27}

\pubyear{2017}

\begin{document}
\label{firstpage}
\pagerange{\pageref{firstpage}--\pageref{lastpage}}
\maketitle

\begin{abstract}
We present high precision, model independent, mass and radius measurements for
16 white dwarfs in detached eclipsing binaries and combine these with previously
published data to test the theoretical white dwarf mass-radius
relationship. We reach a mean precision of 2.4 per cent in mass and 2.7 per
cent in radius, with our best measurements reaching a precision of 0.3 per
cent in mass and 0.5 per cent in radius. We find excellent agreement between
the measured and predicted radii across a wide range of masses and
temperatures. We also find the radii of all white dwarfs with masses less
  than 0.48{\MSUN} to be fully consistent with helium core models, but they
  are on average 9 per cent larger than those of carbon-oxygen core models. In
  contrast, white dwarfs with masses larger than 0.52{\MSUN} all have radii
  consistent with carbon-oxygen core models. Moreover, we find that all but
one of the white dwarfs in our sample have radii consistent with possessing
thick surface hydrogen envelopes ($10^{-5} \ge M_\mathrm{H}/M_\mathrm{WD} \ge
10^{-4}$), implying that the surface hydrogen layers of these white dwarfs are
not obviously affected by common envelope evolution.
\end{abstract}


\section{Introduction}

The overwhelming majority of all stars born in the Galaxy will one day evolve
into white dwarfs. White dwarfs are supported against collapse by electron
degeneracy pressure and as such show the remarkable property that the more
massive they are, the smaller their radius. Moreover, this mass-radius
relationship sets an upper limit to the mass of a white dwarf
\citep{chandrasekhar31} above which electron degeneracy can no longer support
them, a result that underpins our understanding of type Ia supernovae and
hence the expansion of the Universe \citep{riess98,perlmutter99}. The
mass-radius relationship forms an essential part of many studies of
white dwarfs, such as the initial-final mass relationship
\citep[e.g.][]{catalan08}, the white dwarf luminosity function
\citep[e.g.][]{garcia16} as well as allowing us to compute masses from
spectroscopic data alone \citep[e.g.][]{bergeron92,bergeron01}, and underlies
the best mass measurements for white dwarfs in interacting binary systems
\citep[e.g.][]{wood85,littlefair08,savoury11}. 

Despite its huge importance to a wide range of astrophysical topics, the white
dwarf mass-radius relationship remains poorly tested observationally. Previous
efforts involve using astrometric binaries where a white dwarf is paired with
a bright main-sequence star and which have accurate distance measurements and
dynamical masses from orbital fits. These can be combined with spectroscopic
measurements of the gravitational redshift of the white dwarf to yield masses
and radii \citep{shipman97,barstow05,bond15}. 
Unfortunately, very few of these systems have accurate orbital fits and so
masses and radii are instead determined from a combination of parallax and
spectroscopic data \citep[e.g.][]{holberg12}. However, these measurements
still rely on mass-radius relationships to a small extent, since they require
an estimate of the intrinsic flux of the white dwarf (i.e. the monochromatic
Eddington flux), made by fitting the spectrum of the white dwarf with
model atmosphere codes (that in turn rely on a mass-radius relationship). This
semi-empirical method can also be applied to isolated white dwarfs, provided
they have accurate parallax measurements. Therefore, the large sample of 
parallax measurements from the {\it Gaia} mission should allow the detection
of any offsets between observed and theoretical mass-radius relations. However,
due to uncertainties in model atmospheres and evolutionary models, these
results may be difficult to interpret and hence genuine model-independent mass
and radius measurements are still required to properly test mass-radius
relations and fully understand {\it Gaia} data \citep{tremblay17}.

Using isolated white dwarfs, or those in wide astrometric binaries, limits the
range over which one can test the white dwarf mass-radius relationship. This
is because all of these white dwarfs will have masses larger than
$\sim$0.5{\MSUN}, since lower mass white dwarfs are produced from
main-sequence stars with masses less than $\sim$0.8{\MSUN} and the Universe
is not old enough for these stars to have evolved off the main-sequence
yet. However, white dwarfs with masses less than 0.5{\MSUN} have been
identified, and are almost exclusively found in close binary systems, the low
masses being the result of past interactions between the two stars
\citep{marsh95,rebassa11}, although note that massive white dwarfs can also be
found in close binaries \citep[see the latest catalogue from][for
example]{rebassa16}. Moreover, a fraction of these close binaries are
eclipsing, allowing us to directly measure the mass and radius of the white
dwarf to a precision of $1-2$ per cent, independent of model atmosphere
calculations. Double-lined eclipsing binaries are generally the best sources
for mass-radius measurements \citep{torres10}.

White dwarfs with masses below 0.5{\MSUN} are expected to have cores
composed primarily of helium, since they are the result of strong mass-loss
episodes during the red giant branch stage (e.g. from binary interactions),
before the helium core flash has converted the core to carbon and
oxygen. However, it may still be possible to form carbon-oxygen (C/O) core
white dwarfs with masses as low as 0.33{\MSUN} via anomalous mass-loss
episodes on the red giant branch or the core He-burning phase
\citep{prada09}, meaning that both C/O and He core white dwarfs may exist in
this mass range. Since He core white dwarfs are more expanded than those with
C/O cores of the same mass and temperature \citep[e.g.][]{panei07}, precise
enough radius measurements can distinguish between the two possible core
compositions.

In this paper we present precise mass and radius measurements for 16 white
dwarfs in detached eclipsing binaries with low-mass main-sequence star
companions and combine them with 10 previous measurements to test theoretical
mass-radius relationships over a wide range of masses and
temperatures. Results for the main-sequence stars will be presented elsewhere.

\section{Observations and their reduction}

In this section we describe our photometric and spectroscopic
observations. Due to the large number of entries, the observing logs can be
found in Table~\ref{tab:photlog} (photometry) and Table~\ref{tab:speclog}
(spectroscopy) in the appendix. 

\subsection{ULTRACAM photometry}

We obtained high-speed photometry for the majority of our targets using the
frame-transfer CCD camera ULTRACAM \citep{dhillon07}. Observations were
performed between 2002 and 2016 and were obtained with ULTRACAM mounted as a
visitor instrument on the 3.5-m New Technology Telescope (NTT) on La Silla,
Chile, the 4.2-m William Herschel Telescope (WHT) on La Palma, Spain and the
8.2-m Very Large Telescope (VLT) at Paranal, Chile. ULTRACAM uses a triple
beam setup allowing one to obtain data in the $u'$, $g'$ and either $r'$ or
$i'$ band simultaneously.

All of these data were reduced using the ULTRACAM pipeline
software. The source flux was determined with aperture photometry using a
variable aperture scaled according to the full width at half maximum
(FWHM). Variations in observing conditions were accounted for by determining
the flux relative to a comparison star in the field of view. The data were
flux calibrated using observations of standard stars observed during twilight. 

\subsection{ULTRASPEC photometry}

We observed a number of eclipsing systems with the high-speed camera ULTRASPEC
\citep{dhillon14} mounted on the 2.4-m Thai National Telescope (TNT) on Doi
Inthanon, Thailand. ULTRASPEC is a high-speed frame-transfer EMCCD camera that
operates in a similar fashion to ULTRACAM but with a single beam. Our
observations were made using either the $g'$ band filter or a broad $u'+g'+r'$
filter known as $KG5$ (as described in \citealt{dhillon14}, see also the
appendix of \citealt{hardy17}). The ULTRASPEC data were reduced using the
ULTRACAM pipeline as previously described.

\subsection{X-shooter spectroscopy}

\begin{figure*}
  \begin{center}
    \includegraphics[width=0.95\textwidth]{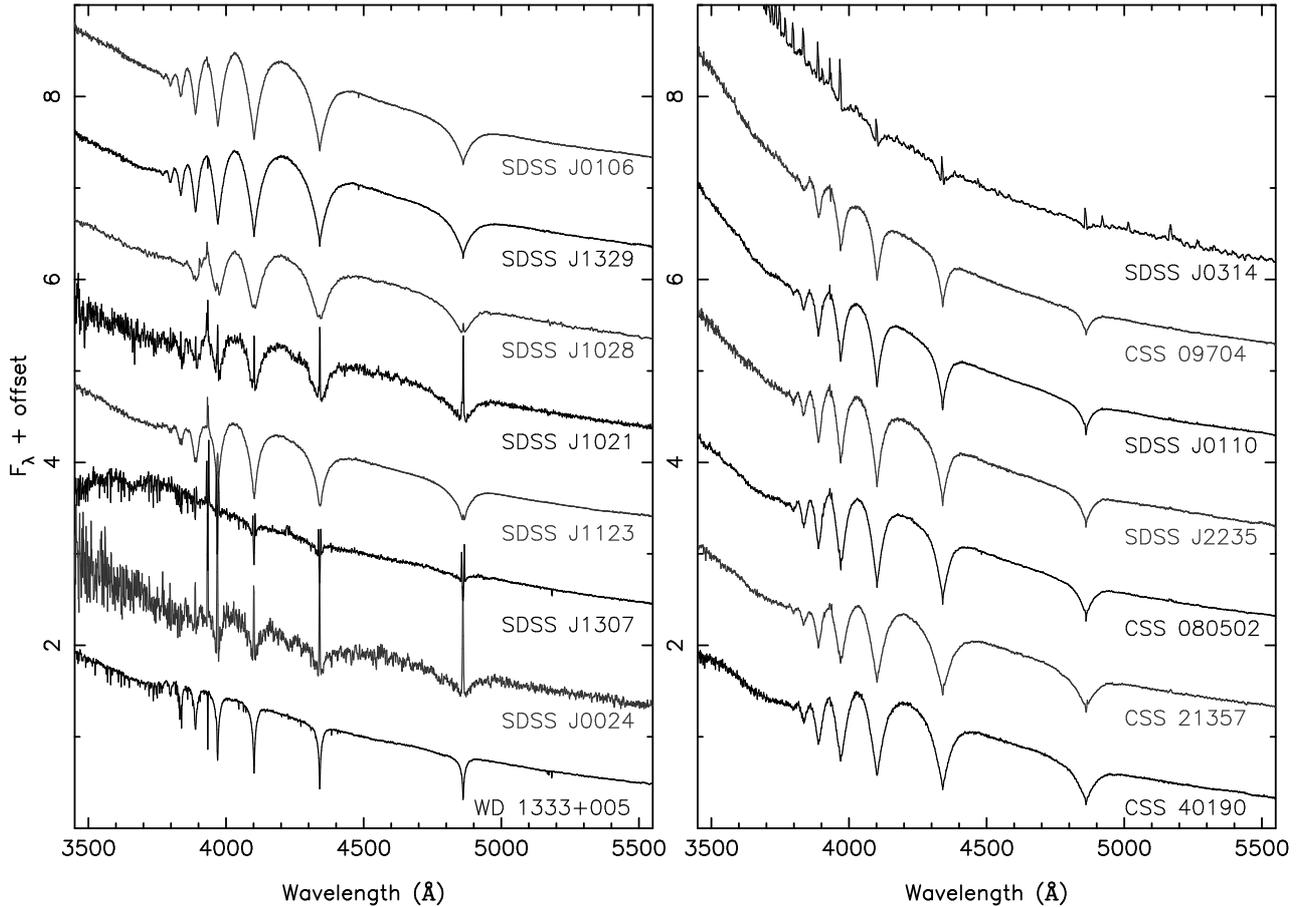}
    \caption{X-shooter UVB arm spectra of the white dwarfs in our binaries
      ordered from coolest (bottom-left) to hottest (top-right), in the white
      dwarf restframes. The M dwarf components have been subtracted from
      these spectra, though any strong, variable emission lines from them
      (e.g. from irradiation or activity) are not completely removed
      (particularly strong in the case of SDSS\,J0314+0206). The spectra have
      been binned by a factor of 10 for clarity.}
  \label{fig:specs}
  \end{center}
\end{figure*}

We spectroscopically observed our eclipsing systems with the medium resolution
echelle spectrograph X-shooter \citep{vernet11}, which is mounted at the
Cassegrain focus of the VLT-UT2 at Paranal, Chile. X-shooter covers the
spectral range from the atmospheric cutoff in the UV to the near-infrared K
band in three separate arms, known as the UVB (0.30$-$0.56 microns), VIS
(0.56$-$1.01 microns) and NIR (1.01$-$2.40 microns). Separate slit widths can
be set for each arm and the UVB and VIS arms can be binned by up to a factor of
$2\times2$. All our observations were performed with slit widths of 1.0'',
0.9'' and 0.9'' in the UVB, VIS and NIR arms respectively. During X-shooter
exposures the evolution of the parallactic angle is not followed. Prior to
2012 the atmospheric dispersion corrector (ADC) allowed observations to be
taken in long series without needing to reorient the slit. The ADCs were
disabled in August 2012 and hence subsequent observations had the potential to
suffer from large slit losses, particularly for a long series of
exposures. This effect was mitigated by setting the slit angle to pass through
the parallactic halfway through a 1 hour set of exposures, re-aligning the
slit after each hour, to ensure that it never drifted too far from the
parallactic. In addition, for all observations the VIS arm was binned by a
factor of 2 in the spatial direction, while the UVB arm was also binned by a
factor of 2 in the spatial direction (for all observations) and by a factor of
2 in the dispersion direction just for our faintest targets ($g>19$). This
results in a resolution of R$\sim$7500.

In addition to our main targets, we also observed a number of M dwarf template
stars with the same instrumental setup (from M1.0 to M8.0 in steps of 0.5 of a
spectral class) as well as several bright DC white dwarfs, which we used to
remove telluric features from our target spectra. All of the data were reduced
using the standard pipeline release of the X-shooter Common Pipeline Library
(CPL) recipes (version 2.6.8) within ESORex, the ESO Recipe Execution Tool. 

For a small number of our targets the NIR arm data had very low
signal-to-noise ratios due to the faintness of these targets. These are
indicated in Table~\ref{tab:speclog} and we discarded these data from our
subsequent analysis. Figure~\ref{fig:specs} shows UVB arm spectra for all of
our targets with the M dwarf components removed (using in-eclipse or
appropriately scaled template spectra, see Section~\ref{sec:specfit}).

The accuracy of the wavelength calibration of X-shooter data from the pipeline
reduction is 0.03nm in the UVB, 0.02nm in the VIS and 0.004nm in the NIR arm,
corresponding to a velocity precision of $\sim$1{\kms} at
H$\alpha$. Additionally, there is a further $\sim$0.01nm offset in the
wavelength solution due to imperfect positioning of the target in the
slit\footnote{https://www.eso.org/sci/facilities/paranal/instruments/xshoo
ter/doc/XS\_wlc\_shift\_150615.pdf}. However, the wavelength accuracy can be
improved in the VIS and NIR arms using sky emission and telluric lines,
enabling velocity measurements to an accuracy of $\sim$0.5{\kms}. Since there
are no sky lines in the UVB arm we used any main-sequence star lines visible
in the UVB arm to measure any systemic velocity offsets (relative to the
main-sequence star features in the VIS/NIR arms). In some cases there are no
such features visible, in which case we consider the accuracy of any velocity
measurements in the UVB arm to be 1\kms. We find no evidence of variations in
the wavelength shifts between subsequent spectra of the same target,
therefore, this primarily affects the precision of our systemic velocity
measurements, rather than the radial velocity amplitudes.

\section{Breaking the degeneracy between inclination and scaled radii} \label{sec:degen}

Observations of just the eclipse of the white dwarf by its main-sequence star
companion do not contain enough information to fully solve for the binary
parameters. This is because the eclipse profile only contains two pieces of
information (its width and the duration of the ingress/egress), while
there are three unknowns: the orbital inclination, $i$, and the radii of the
two stars scaled by the orbital separation ($R_\mathrm{WD}/a$ and
$R_\mathrm{MS}/a$). Therefore, it is possible to fit the same eclipse profile
with a high inclination system containing a large white dwarf and small
main-sequence star or a lower inclination system with a smaller white dwarf
and larger main-sequence star. Hence at least one more piece of information is
required in order to break this degeneracy. In this section we outline several
different techniques that we use to this end.  

\subsection{The depth of the secondary eclipse}

The most straightforward and direct method of breaking the degeneracy between
the inclination and scaled radii is to measure the depth of the secondary
eclipse (i.e. the transit of the white dwarf across the face of its
main-sequence star companion). This is analogous to an exoplanet transit and
hence the depth of the eclipse is the ratio of the areas of the two stars,
$(R_\mathrm{WD}/R_\mathrm{MS})^2$. Therefore, combining this with the primary
eclipse profile directly yields the inclination and two radii \citep[see][for
an example of this technique]{parsons10}.

However, since white dwarfs are so much smaller than main-sequence stars and
because the white dwarf dilutes the eclipse, the depth of the secondary eclipse
is usually less than 1-2 per cent. Given the relative faintness of these
objects (the mean $g$ band magnitude of our sample is 17.6), in the vast
majority of systems, detecting and measuring the depth of the secondary eclipse
is currently not possible and other techniques are required. The secondary
eclipse is only detected in one of our systems, RR\,Cae.

\subsection{The gravitational redshift of the white dwarf}

\begin{figure}
  \begin{center}
    \includegraphics[width=0.95\columnwidth]{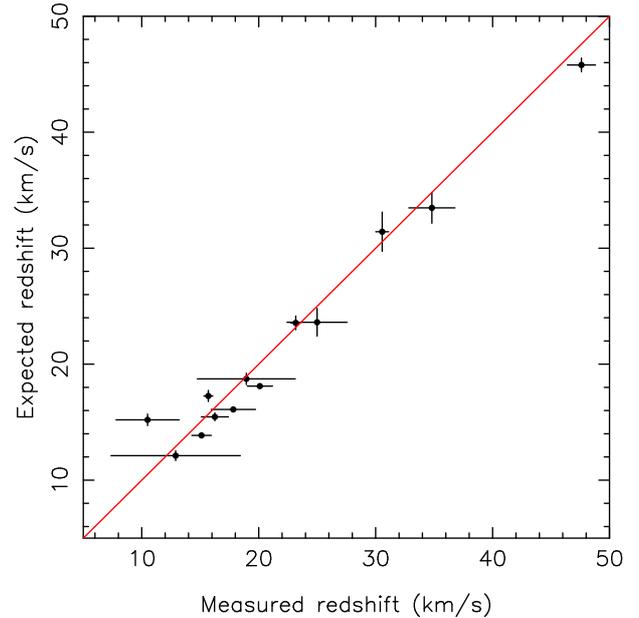}
    \caption{The measured and computed gravitational redshifts of the white
      dwarfs in our binaries. The expected redshift values are those
      determined from the mass and radius of the white dwarf, computed from
      Equation~\ref{eqn:red}. The measured values are the spectroscopic
      redshifts from our X-shooter data. Only objects where the spectroscopic
      redshift was not used to determine the inclination are shown.} 
  \label{fig:shift}
  \end{center}
\end{figure}

General relativity tells us that the light emerging from the gravitational
well of a white dwarf is redshifted by an amount given by 
\begin{equation}
V_z = 0.635 (M_\mathrm{WD}/M_{\sun})(R_{\sun}/R_\mathrm{WD})\, \kms. \label{eqn:red}
\end{equation}
Furthermore, if we know the radial velocity amplitudes of the two stars then
Kepler's third law tells us
\begin{equation}
M_\mathrm{WD} = \frac{P K_\mathrm{MS} (K_\mathrm{WD}+K_\mathrm{MS})^2}{2 \pi G \sin^3{i}},
\end{equation}
where $P$ is the orbital period and $K_\mathrm{WD}$ and $K_\mathrm{MS}$ are
the radial velocity semi-amplitudes of the white dwarf and main-sequence star
respectively. Therefore, if we can spectroscopically measure the gravitational
redshift of the white dwarf (measured from the offset in the radial velocity
semi-amplitudes of the two stars), then combining these two equations gives us a
relationship between the white dwarf's radius and the binary inclination. This
can then be used in combination with the primary eclipse profile to break the
degeneracy between the inclination and scaled radii of the two
stars. See \citet{parsons12} for an example of this technique.

The advantage of this technique is that all it requires is that spectral
features from both stars are visible. Therefore, it is applicable to a wide
range of systems. However, the precision of the final parameters is strongly
correlated to how well the radial velocity semi-amplitudes of the two stars
can be measured and therefore it can be less precise than some of the other
methods, particularly in systems where one star strongly dominates over the
other in the spectrum.

Finally, we note that there has been some discrepancy between the measured
and expected gravitational redshift of the white dwarf in the wide binary
Sirius \citep{barstow05} and several previous redshift measurements of white
dwarfs in close binaries were only marginally in agreement with the expected
value \citep{maxted07,parsons10,pyrzas12}. Additionally, in several cases, we
use features from the white dwarf that originate from accreting material from
the wind of the main-sequence star (and may form higher up in the white
dwarf's atmosphere, hence lower gravitational potential, particularly so in
the case of emission lines from the white dwarf). Therefore, it is worth
checking the accuracy of this method. In Figure~\ref{fig:shift} we show the
measured and expected redshift values for all of the white dwarfs in our
sample (and previously published) in which the gravitational redshift itself
was not used to determine the stellar parameters (i.e. the method outlined
above was not used). The values are in excellent agreement, even when using
lines originating from accreting wind material. Therefore, stellar parameters
determined from this technique should be accurate and also shows that our
final mass and radius constraints are consistent regardless of the method used
to break the inclination degeneracy.

\subsection{The rotational broadening of the M star} \label{sec:rbr}

The short periods and extreme mass ratios of these binaries means that the
main-sequence star components are tidally locked to the white dwarf and
therefore their rotational period matches the orbital period (for low-mass M
stars in binaries with a white dwarf and periods of less than a day, the tidal
synchronisation timescale is of the order of $10^5$ years and so all of our
systems should be synchronised \citealt{zahn77}). This means that these stars
are rotating quite rapidly and hence their spectral lines are broadened by a
factor given by 
\begin{equation}
V_\mathrm{rot}\sin{i} = K_\mathrm{MS}(1+q)\frac{R_\mathrm{MS}}{a},
\end{equation}
where $q=M_\mathrm{WD}/M_\mathrm{MS}$, the binary mass ratio
\citep{marsh94}. The rotational broadening can be measured from spectroscopy
by artificially broadening the lines of template stars to fit the observed
line profiles of our systems, taking into account any additional smearing of
the lines from the velocity shift of the main-sequence star during an exposure
\citep[see][for details of this method]{marsh94}. Note that we applied a
high-pass filter to both the observed and broadened template spectra before
comparing them in order to prevent the continuum dominating the rotational
broadening determination. If both $K_\mathrm{MS}$ and $q$ have also been
spectroscopically measured, then the rotational broadening gives a direct
measurement of the scaled radius of the main-sequence star, which can be
combined with the primary eclipse profile to fully determine the binary and
stellar parameters \citep[see][for an example of this technique]{parsons16}.

This technique is useful for systems in which the main-sequence star
contributes a substantial fraction of the optical flux. Moreover, it is best
suited to systems with very short orbital periods and those that are close to
Roche-lobe filling, where the rotational broadening is maximised. However, in
systems where the white dwarf completely dominates the optical flux, or where
the main-sequence star is strongly irradiated by the white dwarf, this
technique may not be reliable or even possible.

\subsection{The amplitude of ellipsoidal modulation}

The tidal distortion of main-sequence stars in close binaries with white
dwarfs causes a sinusoidal variation in the light curve on half the binary
period, caused by the variation in surface area that the star presents to the
Earth during its orbit. The amplitude of this effect is approximately given by
\begin{equation}
\frac{\delta F}{F} = 0.15 \frac{(15 + u_\mathrm{MS})(1 + \beta_\mathrm{MS})}{3 - u_\mathrm{MS}} \left(
  \frac{R_\mathrm{MS}}{a} \right)^3 q \sin^2{i},
\end{equation}
where $u_\mathrm{MS}$ is the linear limb darkening coefficient and
$\beta_\mathrm{MS}$ is the gravity darkening exponent of the main-sequence
star \citep{morris93,zucker07}; note that the companion must be tidally locked
for this assumption to be valid. Therefore, this effect can be used to
establish a relationship between the scaled radius of the main-sequence star
and the inclination in a manner independent of the primary eclipse profile,
and hence can be combined to break the degeneracy between these parameters
\citep[see][for an example of this technique]{parsons12uc}.

However, while this technique can work in some systems, it is often difficult
to reliably measure the ellipsoidal amplitude due to the unknown contribution
from starspots, which may dilute or strengthen the ellipsoidal amplitude
depending upon their location on the surface of the star. Therefore, in this
paper we do not use this technique to help break the degeneracy between the
scaled radii and inclination. However, when visible in the light curve, we do
check that the amplitude is consistent with our best-fit models.

\section{Radial velocity measurements} \label{sec:rvs}

\begin{figure*}
  \begin{center}
    \includegraphics[width=0.95\textwidth]{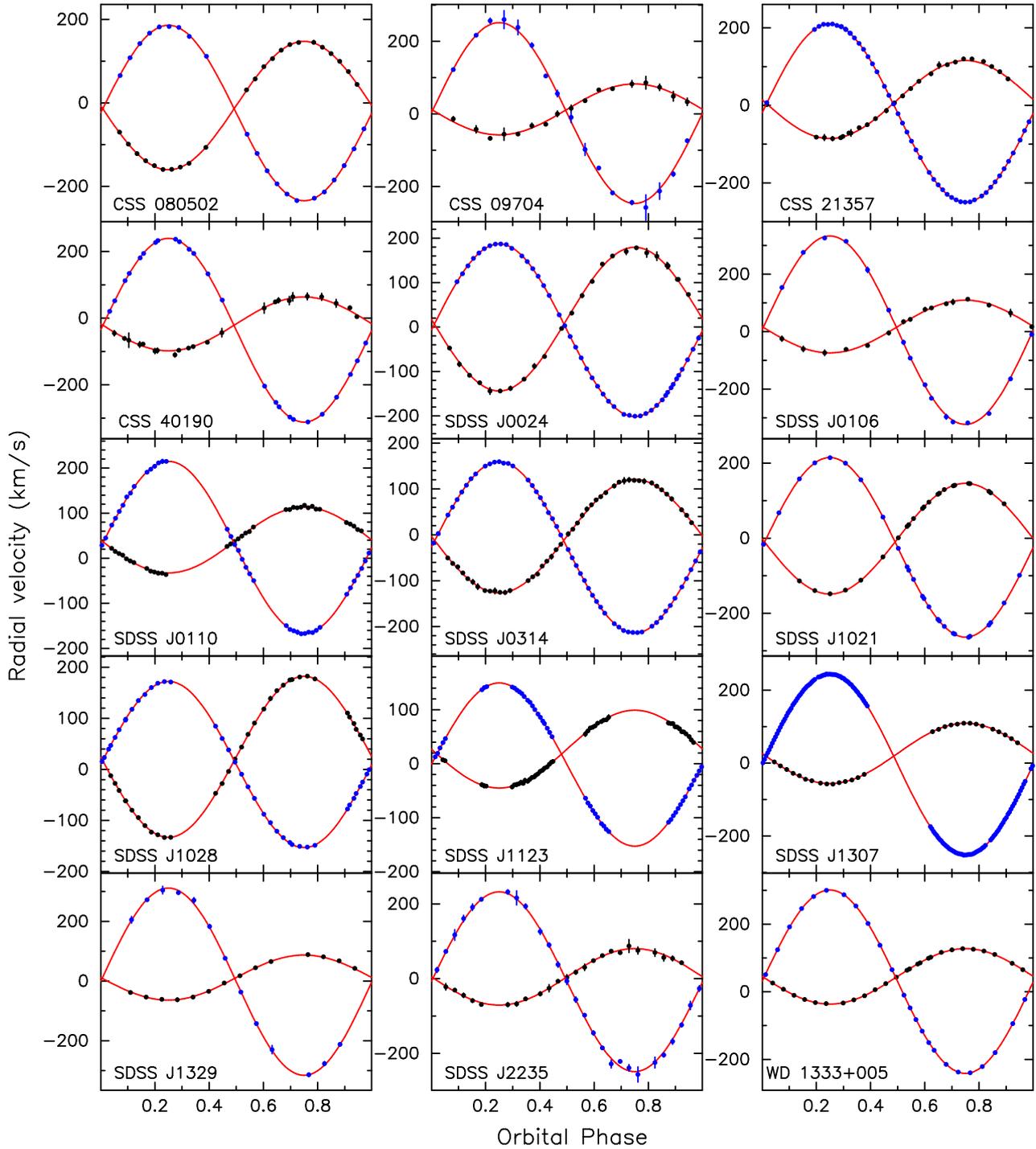}
    \caption{Radial velocity measurements for the white dwarfs (black) and
      main-sequence stars (blue) for all the binaries we observed with
      X-shooter (note that RR\,Cae was not observed with X-shooter). The best
      fit curves are shown in red.}
  \label{fig:rvcurve}
  \end{center}
\end{figure*}

\begin{table*}
 \centering
  \caption{Spectroscopic measurements. We have highlighted with an asterisk 
  ($^*$) those systems where the radial velocity of the main-sequence star was
  determined by applying a correction factor to the velocity of emission lines
  (see Section~\ref{sec:rvs}). Data for RR\,Cae were taken from
  \citet{ribeiro13}. Note that when referring to the hydrogen absorption lines
  of white dwarfs we actually fitted the narrow core of the line, rather than
  the broader wings.}
  \label{tab:specmeas}
  \begin{tabular}{@{}lccccccc@{}}
  \hline
  Object         & $K_\mathrm{WD}$ & $\gamma_\mathrm{WD}$ & WD feature & $K_\mathrm{MS}$ &  $\gamma_\mathrm{MS}$ & MS feature & $V_\mathrm{rot}\sin{i}$ \\
                 &   ($\kms$)    &         ($\kms$)    & used       &   ($\kms$)    &         ($\kms$)    & used        & ($\kms$)              \\
  \hline
CSS\,080502      & $153.89\pm0.99$ & $-6.51\pm0.65$  & H$\beta$          & $210.33\pm0.43$ & $-24.34\pm1.79$ & \Ion{Na}{i} 8183/94 & $131.76\pm2.56$ \\
CSS\,09704       & $70.2\pm5.1$    & $12.5\pm4.9$    & H$\alpha$         & $249.1\pm7.4^*$ & $2.5\pm6.8$     & H$\alpha$ (em)      & -               \\
CSS\,21357       & $100.63\pm2.51$ & $15.06\pm1.96$  & H$\beta$          & $230.27\pm0.50$ & $-19.75\pm0.50$ & \Ion{Na}{i} 8183/94 & $60.78\pm2.06$  \\
CSS\,40190       & $80.85\pm4.74$  & $-17.09\pm4.09$ & H$\beta$          & $275.92\pm0.88$ & $-36.03\pm0.92$ & \Ion{Na}{i} 8183/94 & $74.75\pm1.45$  \\
RR\,Cae          & $73.3\pm0.4$    & $98.0\pm1.0$    & H$\beta$          & $195.1\pm0.3$   & $77.9\pm0.4$    & H$\alpha$           & -               \\
SDSS\,J0024+1745 & $161.67\pm3.17$ & $17.82\pm2.45$  & \Ion{Ca}{ii} 3934 & $194.30\pm0.20$ & $-7.17\pm0.77$  & \Ion{Na}{i} 8183/94 & $107.42\pm1.80$ \\
SDSS\,J0106$-$0014 & $91.86\pm3.49$  & $17.83\pm3.44$  & H$\alpha$         & $328.15\pm3.22$ & $5.04\pm3.16$   & \Ion{Na}{i} 8183/94 & -               \\
SDSS\,J0110+1326 & $72.90\pm1.29$  & $40.36\pm0.78$  & H$\alpha$         & $190.85\pm1.23$ & $24.11\pm0.86$  & \Ion{K}{i} 12522    & $74.20\pm4.40$  \\
SDSS\,J0314+0206 & $123.69\pm2.04$ & $-2.81\pm1.35$  & \Ion{He}{i} 5876  & $186.77\pm0.50$  & $-27.31\pm0.58$ & \Ion{K}{i} 12522    & -               \\
SDSS\,J1021+1744 & $147.45\pm0.60$ & $-1.20\pm0.50$  & H$\alpha$ (em)    & $239.21\pm0.70$ & $-24.37\pm0.50$ & \Ion{Na}{i} 8183/94 & $140.56\pm2.26$ \\
SDSS\,J1028+0931 & $157.75\pm1.00$ & $24.64\pm0.80$  & H$\alpha$ (em)    & $162.68\pm0.50$ & $9.53\pm0.50$   & \Ion{Na}{i} 8183/94 & $90.38\pm1.50$  \\
SDSS\,J1123$-$1155 & $72.09\pm1.39$  & $27.07\pm0.90$  & \Ion{Ca}{ii} 3934 & $151.17\pm0.20$ & $-1.44\pm0.50$  & \Ion{Na}{i} 8183/94 & -               \\
SDSS\,J1307+2156 & $82.80\pm0.59$  & $26.59\pm0.50$  & \Ion{Ca}{ii} 3934 & $247.89\pm0.28$ & $-3.97\pm0.50$  & \Ion{K}{i} 12522    & $54.56\pm2.56$  \\
SDSS\,J1329+1230 & $75.50\pm1.70$  & $11.72\pm1.87$  & \Ion{Ca}{ii} 3934 & $313.9\pm5.8^*$ & $-2.03\pm1.20$  & H$\alpha$ (em)      & -               \\
SDSS\,J2235+1428 & $75.52\pm4.74$  & $4.75\pm3.43$   & H$\alpha$         & $240.56\pm4.64$ & $-8.15\pm4.34$  & \Ion{Na}{i} 8183/94 & $65.86\pm8.50$  \\
WD\,1333+005     & $81.95\pm0.47$  & $45.67\pm0.50$  & \Ion{Ca}{ii} 3934 & $271.19\pm0.20$ & $29.98\pm0.50$  & \Ion{Na}{i} 8183/94 & $70.56\pm2.06$  \\
  \hline
\end{tabular}
\end{table*}

We measured the radial velocity semi-amplitudes of both stars in each binary
by first identifying the cleanest features from each star then fitting them
with a combination of a straight line and a Gaussian (or double Gaussian in
the case of Balmer absorption from the white dwarf, that includes a broader
component for the wings of the line and a narrower one for the core of the
line). Each spectrum was fitted individually and the resulting velocity
measurements ($v_t$) were combined to determine the semi-amplitude ($K$) and
systemic velocity ($\gamma$) of each star by fitting them with the following
equation:
\begin{equation}
v_t = \gamma + K \sin{\left(\frac{2 \pi (t-t_0)}{P}\right)},
\end{equation}
where $P$ is the orbital period and $t_0$ is the mid-eclipse time (previously
determined from the photometry). The results for all systems are listed in
Table~\ref{tab:specmeas}, which also gives the specific features used to
measure the velocities of each star, as well as the rotational broadening
(where possible) for the main-sequence stars measured as outlined
in Section~\ref{sec:rbr}. Recall that the difference in the $\gamma$
  values between the white dwarf and main-sequence star is related to the
  gravitational redshift of the white dwarf. We also show the radial velocity
curves for all our X-shooter targets in Figure~\ref{fig:rvcurve} (note that
the system RR\,Cae is not included in this figure, since the radial velocity
data have already been published in \citealt{ribeiro13}).

For some of our objects, specifically those containing very hot white dwarfs,
the main-sequence star is irradiated to such an extent that its spectral
features are completely diluted. In these cases, many absorption lines from
the main-sequence star are instead in emission. However, while these emission
lines can be strong and hence give very precise velocity measurements, because
they arise from the heated face of the star they do not track the
centre-of-mass of the star, but rather the centre-of-light of the emission
region on the hemisphere facing the white dwarf. Therefore, they are not ideal
features to use for precise stellar parameter measurements. However, in some
cases these emission lines are the only features visible from the
main-sequence star. It is possible to correct from the emission line velocity
($K_\mathrm{emis}$) to the centre-of-mass velocity via an equation,
\begin{equation} \label{eqn:kcorr}
K_\mathrm{MS} = \frac{K_\mathrm{emis}}{1-f(1+q)\frac{R_\mathrm{MS}}{a}},
\end{equation}
where $f$ is a constant between 0 and 1 which depends upon the location of the
centre of light \citep{parsons12sp}. For $f=0$ the emission is spread
uniformly across the entire surface of the secondary star and therefore the
centre of light is the same as the centre of mass. For $f=1$ all of the flux
is assumed to come from the point on the main-sequence star's surface closest
to the white dwarf (the substellar point). This sets the upper and lower
limits on the true centre-of-mass velocity. However, we can be more precise
than this, since this $f$ factor is related to the optical depth of the
emission \citep{parsons10}. With a sufficient number of emission lines from
different atomic species it is possible to reliably estimate the
centre-of-mass radial velocity to within $\sim5\kms$. This technique was
tested by \citet{parsons12} for SDSS\,J1212$-$0123, where both absorption and
emission features are present.

Finally, there are some systems where the irradiation is fairly mild and so
absorption lines from the main-sequence star are still visible, but there is a
subtle emission core to the line that slightly shifts the velocity of the
line. In these cases it is important to test if this is having a noticeable
effect by measuring the equivalent width of the line as a function of orbital
phase. If the line is weaker when the heated side of the star is viewed face
on then the velocity will not be reliable. We did this for all the lines we
used to measure the radial velocity of the main-sequence star and found that
they were clean and reliable (i.e. showed no variations in their
  equivalent widths throughout the orbit).

\section{Spectral fitting}

In this section we detail how we removed the main-sequence star features from
our spectra and how the resulting white dwarf spectra were then fitted to
determine their effective temperatures. 

\subsection{Main-sequence star removal} \label{sec:specfit}

In order to accurately fit the spectra of the white dwarfs in our binaries we
need to remove the main-sequence star contributions. This is important since
fitting spectra with both white dwarf and main-sequence star models
simultaneously has been shown to give erroneous results when the main-sequence
star is particularly dominant \citep{parsons13} and reliable white dwarf
effective temperature measurements are essential so that we can compare our
mass and radius measurements with the correct models.

We spectroscopically observed all of our X-shooter targets during the eclipse
of the white dwarf in order to obtain clean spectra of just the main-sequence
star components. For the majority of our targets this was successful, however,
for four of our objects (CSS\,09704, SDSS\,J0106$-$0014, SDSS\,J1329+1230 and
SDSS\,J2235+1428) the in-eclipse spectra had extremely low signal-to-noise
ratios (S/N$<$1) due to a combination of the faintness of the main-sequence
components and the short duration of the eclipse limiting the exposure
time. We then used these in-eclipse spectra (or template spectra in the case
of the four low S/N objects, determined from their measured masses and the
mass-spectral class relationship from \citealt{baraffe96}) and subtracted them
from all other spectra after shifting them into the main-sequence star's rest
frame. We then shifted these main-sequence star subtracted spectra into the
rest frame of the white dwarf and averaged them to create the spectra shown in
Figure~\ref{fig:specs}. 

While this process removes the absorption features of the main-sequence stars,
the majority of emission lines were not removed, since these tend to be highly
variable as they arise from a combination of irradiation and stellar
activity, hence these are still present in the final spectra. For example, the
highly irradiated main-sequence star in SDSS\,J0314+0206 produces a large
number of emission lines as seen in Figure~\ref{fig:specs}. Furthermore, in
systems where the main-sequence star is highly irradiated, or very close to
filling its Roche lobe, the absorption features can vary in strength with
orbital phase due to temperature variations over the surface of the star,
hence the subtraction is never perfect and we therefore limit our spectral
analysis to the shortest wavelength data (the UVB arm of X-shooter) where the
white dwarfs generally dominate the flux anyway, to reduce this effect as much
as possible.

\subsection{White dwarf spectral fitting}

The temperature of a white dwarf can have a significant effect on its radius,
in particular for relatively hot, low-mass white dwarfs. We fitted the
X-Shooter spectra of the white dwarfs in our sample following the method
outlined in \citet{rebassa07}. The X-shooter spectra of each white dwarf were
shifted into its rest-frame, and had the contribution of the M-dwarf removed
(see Section~\ref{sec:specfit} and Figure~\ref{fig:specs}). We used an updated
model grid of hydrogen (DA) atmosphere white dwarfs computed with the code
described in \citet{koester10}, adopting $\mathrm{ML2/}\alpha=0.8$ for the
mixing length in convective atmospheres ($T_\mathrm{eff}\la15,000$\,K). The
continuum-normalised Balmer lines in the X-Shooter spectra were fitted with
the full model grid, and the best-fit parameters were then determined from a
bicubic fit to the $\chi^2$ surface in the $(T_\mathrm{eff}, \log g)$
plane. The degeneracy between the ``hot'' and ``cold'' solution, corresponding
to similar equivalent widths of the Balmer lines, was resolved using the slope
of the white dwarf continuum.

To test the robustness of the fits, and estimate realistic parameters, we
fitted both the average and all individual X-Shooter spectra. The average
spectra were of high signal-to-noise ratio ($\mathrm{S/N}\simeq60-140$, apart
for SDSS\,J0024+1745 ($\mathrm{S/N}\simeq12$) and SDSS\,J1021+1744
($\mathrm{S/N}\simeq40$)), and uncertainties in the fits to these spectra are
therefore dominated by systematic effects. In contrast, the fits to the
individual X-Shooter spectra (typical $\mathrm{S/N}\simeq10-20$) are dominated
by the statistical uncertainties. We compute the best-fit parameters for each
system as the average from the fits to the individual spectra, and adopt as
uncertainties the standard deviations from these mean values. For all systems,
the fit parameters derived from the average spectrum are, within the
uncertainties, consistent with these best-fit values (see \citealt{tremblay17}
for a detailed discussion on the precision of atmospheric parameters measured
from multiple spectra). The X-Shooter spectra (average and individual) of
SDSS\,J1021+1744, SDSS\,J0024+1745, and SDSS\,J0314+0206 were significantly
contaminated by emission lines from the active and/or irradiated companion
star. We masked the regions affected by the emission lines before fitting the
spectra, and the atmospheric parameters of these stars are subject to larger
uncertainties.

In the case of SDSS\,J1307+2156, the Balmer lines were so weak and affected
by emission lines that the fitting procedure failed. Instead we fitted the
average X-Shooter spectrum (this time including the M-dwarf contribution),
supplemented by the SDSS $ugriz$ fluxes and the GALEX near-ultraviolet flux
using a composite spectrum of a DA white dwarf model, with the surface gravity
fixed $\log g=8.062$, as determined from the light curve fits. We found that
the spectral energy distribution of the white dwarf in SDSS\,J1307+2156 is
well reproduced by $T_\mathrm{eff}=8500\pm500$\,K. 

\section{Light curve fitting}

All light curves were fitted using a code written for the general case
of binaries containing white dwarfs (see \citealt{copperwheat10} for a
detailed description). The program subdivides each star into small elements
with a geometry fixed by its radius as measured along the direction of centres
towards the other star. Roche geometry distortion and irradiation of the
main-sequence star are included.

The basic parameters required to define the model are:
\begin{itemize}
\item $q = M_\mathrm{MS}/M_\mathrm{WD}$, the binary mass ratio
\item $V_s = (K_\mathrm{WD}+K_\mathrm{MS})/\sin{i}$, the sum of the
  unprojected stellar orbital speeds
\item $i$, the binary inclination
\item $R_\mathrm{WD}/a$ and $R_\mathrm{MS}/a$, the radii of the two stars
  scaled by the orbital separation, $a$.
\item $T_{0}$, the time of mid-eclipse
\item $P$, the binary period
\item $T_\mathrm{eff,WD}$ and $T_\mathrm{eff,MS}$, the unirradiated
  temperatures of the two stars. Note that the temperatures are essentially
  just flux scaling parameters and only approximately correspond to the actual
  temperatures.
\item $A$, the fraction of the irradiating flux from the white dwarf absorbed
  by the main-sequence star 
\item Gravity darkening coefficients for the main-sequence star
\item Limb darkening coefficients for both stars
\end{itemize}

$q$, $V_s$, $i$, $R_\mathrm{WD}/a$, $R_\mathrm{MS}/a$ and $T_{0}$ were allowed
to vary, although any spectroscopic constraints that we had for $V_s$ and $q$
were included in the fitting process. All data were phase-folded using
previously published ephemerides, hence the period was kept fixed at a value
of 1. We kept $T_\mathrm{eff,WD}$ fixed at the value determined from the
spectroscopic fits and allowed $T_\mathrm{eff,MS}$ to vary (we reiterate that
these are only flux scaling parameters and so do not represent the true
temperatures).

The irradiation is approximated by
$\sigma T^{\prime}{}_\mathrm{MS}^{4} = \sigma T_\mathrm{MS}^{4}+ A F_\mathrm{irr}$
where $T^{\prime}{}_\mathrm{MS}$ is the modified temperature and
$T_\mathrm{MS}$ is the temperature of the unirradiated main-sequence star,
$\sigma$ is the Stefan-Boltzmann constant, $A$ is the fraction of the
irradiating flux from the white dwarf absorbed by the main-sequence star and
$F_\mathrm{irr}$ is the irradiating flux, accounting for the angle of
incidence and distance from the white dwarf. The value of $A$ makes little
difference to the shape of the eclipse of the white dwarf, since at this phase
the heated side of the main-sequence star is pointing away from
Earth. Therefore, for most of our targets (those with just white dwarf eclipse
data) this was fixed at a value of 0.5. Otherwise it was allowed to vary
freely. 

The gravity darkening coefficients for the main-sequence stars were taken from
\citet{claret11} for the appropriate filter. Note that this only has an effect
in systems that are very close to Roche lobe filling. For both stars we
adopted a four-parameter non-linear limb darkening model (see
\citealt{claret00}), given by
\begin{equation}
\frac{I(\mu)}{I(1)} = 1 - \displaystyle\sum_{k=1}^{4}a_{k}(1-\mu^{\frac{k}{2}}),
\end{equation}
where $\mu = \cos{\phi}$ ($\phi$ is the angle between the line of sight
and the emergent flux), and $I(1)$ is the monochromatic specific intensity at
the centre of the stellar disk. Parameters for the white dwarfs were taken
from \citet{gianninas13} and from \citet{claret11} for the main-sequence
stars and held fixed during the fitting process. While this means that our
mass-radius measurements are not completely independent of model atmosphere
calculations, the adopted values of the limb darkening coefficients have a
very minor impact on the final physical parameters of the stars, typically at
a level below their statistical uncertainties \citep{parsons16}.

\begin{figure}
  \begin{center}
    \includegraphics[width=\columnwidth]{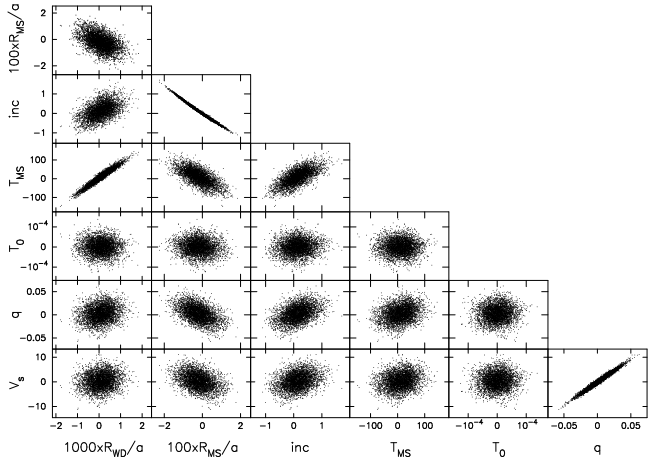}
    \caption{An example of the posterior probability distributions from one of
    our MCMC chains fitting the light curve of SDSS\,J0024+1745. In this case
    the probability was driven by a combination of the $\chi^2$ of the fit to
    the light curve and a prior probability from the spectroscopic mass ratio,
    radial velocity and rotational broadening measurements. The mean values of
    each parameter have been subtracted and only 10 per cent of the points are
    shown for clarity.}
  \label{fig:covar}
  \end{center}
\end{figure}

We used the Markov Chain Monte Carlo (MCMC) method to determine the
distributions of our model parameters \citep{press07}. The MCMC method
involves making random jumps in the model parameters, with new models being
accepted or rejected according to their probability computed as a Bayesian
posterior probability. This probability is driven by a combination of the
$\chi^2$ of the fit to the light curve data and any additional prior
probabilities from our spectroscopic data, which allow us to implement the
inclination and radii degeneracy breaking techniques outlined in
Section~\ref{sec:degen}. For each target an initial MCMC chain was used to
determine the approximate best parameter values and covariances. These were
then used as the starting values for longer chains which were used to
determine the final model values and their uncertainties. Four chains
were run simultaneously to ensure that they converged on the same values.
The first 50,000 points from each chain were classified as a ``burn-in''
  phase and were removed for the subsequent analysis, to ensure the final
  results were not skewed towards the initial values. An example of the
typical parameter distributions is shown in Figure~\ref{fig:covar}. 

\section{Results}

\begin{figure*}
  \begin{center}
    \includegraphics[width=0.9\textwidth]{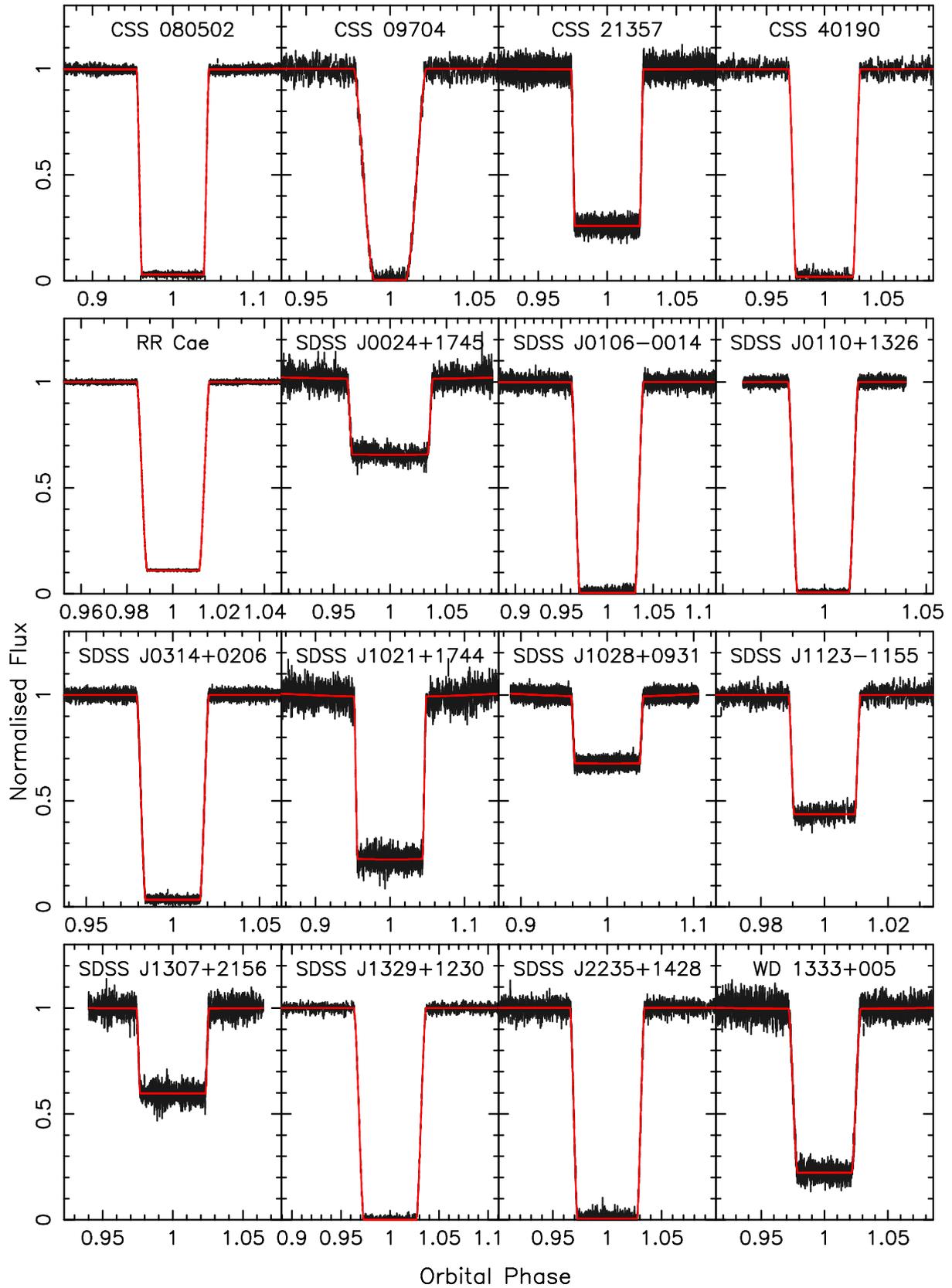}
    \caption{Light curves of the eclipse of the white dwarf in all our
      systems. The best fit models are shown in red. The light curves shown
      are in the $g'$ band or the $KG5$ band (for CSS\,21357,
      SDSS\,J1028+0931, SDSS\,J1123$-$1155, SDSS\,J1307+2156 and WD\,1333+005).} 
  \label{fig:ecl_fit}
  \end{center}
\end{figure*}

\begin{figure*}
  \begin{center}
    \includegraphics[width=0.9\textwidth]{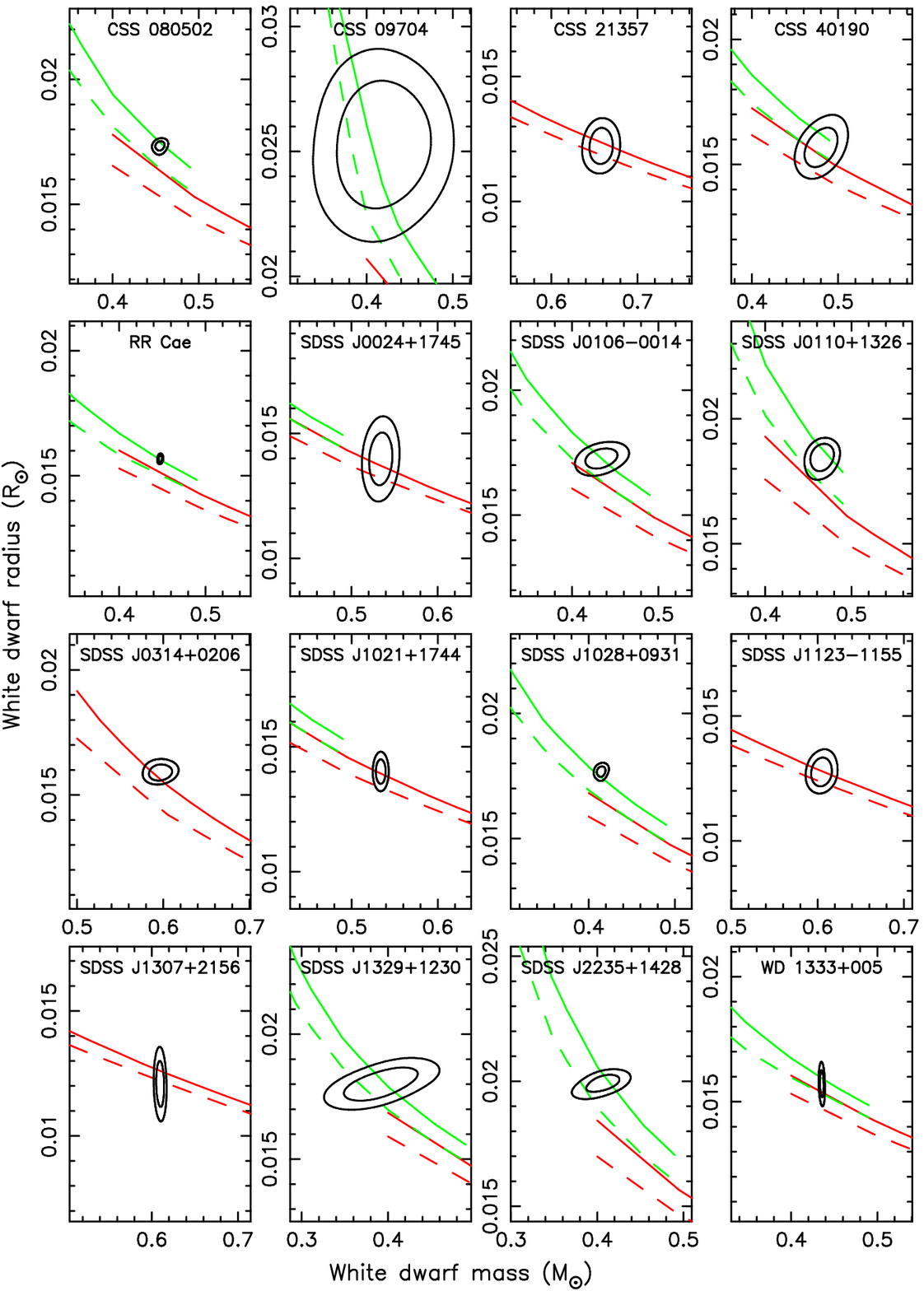}
    \caption{Constraints on the masses and radii of the white dwarfs in all
      our systems, shown as contours (68 and 95 percentile regions). C/O core
      models for the measured white dwarf temperature are shown in red for
      thick (solid, $M_\mathrm{H}/M_\mathrm{WD}=10^{-4}$) and thin (dashed,
      $M_\mathrm{H}/M_\mathrm{WD}=10^{-10}$) hydrogen envelopes
      \citep{fontaine01} and He core models are shown in green for thick
      (solid, $M_\mathrm{H}/M_\mathrm{WD}=10^{-4}$) and thin (dashed,
      $M_\mathrm{H}/M_\mathrm{WD}=10^{-8}$) hydrogen envelopes
      \citep{panei07}. All plots are on the same scale and centred on the mean
      value for each white dwarf.} 
  \label{fig:mr_conts}
  \end{center}
\end{figure*}

In this section we detail the results for each object individually. We
reiterate that the focus of this paper is testing the mass-radius relation for
white dwarfs and the results for the main-sequence stars will be presented in
a forthcoming paper. Model fits to all the primary eclipse light curves are
shown in Figure~\ref{fig:ecl_fit}. Additionally, in Figure~\ref{fig:mr_conts}
we plot the mass-radius constraints for all our white dwarfs along with
different mass-radius relationships corresponding to C/O core models and He
core models with thick and thin surface hydrogen envelopes. White dwarf
  models are taken from \citet{fontaine01} and \citet{benvenuto99}
  ($T_\mathrm{eff}>40,000$\,K) for C/O core white dwarfs and from
  \citet{panei07} for He core white dwarfs. In the following sections we quote
  the mass-radius values and their uncertainties from the 1$\sigma$ contours
  in Figure~\ref{fig:mr_conts}.

\subsection{CSS\,080502}

CSS\,080502 (SDSS\,J090812.04+060421.2 in SIMBAD) was discovered to be an
eclipsing white dwarf plus M dwarf binary with a period of 3.6\,h by
\citet{drake09}. It is a relatively bright ($V=17.1$) system containing a
17,800\,K white dwarf with an M3.5 companion. This binary is ideal for precise
mass-radius measurements since the white dwarf's spectral features dominate
blueward of H$\alpha$ while the M star's features dominate redward from
this. Irradiation of the main-sequence star is also minor. For this system we
used the rotational broadening of the main-sequence star to help constrain the
orbital inclination.

The fit shows that the white dwarf has a mass and radius of
$0.476\pm0.004${\MSUN} and $0.0175\pm0.0003${\RSUN}, placing it in the
region where both C/O and He core white dwarfs may exist. However, while the
radius is consistent with He core models, it is substantially
larger than C/O core models predict and would be 6 per cent oversized if it
has a C/O core (see Figure~\ref{fig:mr_conts}). The radius is also more
consistent with a thick surface hydrogen envelope
($M_\mathrm{H}/M_\mathrm{WD}=10^{-4}$) rather than thinner models.

\subsection{CSS\,09704}

CSS\,09704 (SDSS\,J220823.66-011534.1 in SIMBAD) is a hot (30,000\,K) white
dwarf with a low-mass M6 companion in a 3.8\,h binary discovered by
  \citet{drake10}. The white dwarf completely dominates the optical flux and
substantially irradiates its main-sequence companion star, to the extent that
no absorption lines from the companion are visible in the X-shooter
spectrum. However, several strong emission lines originating from this star
are seen (caused by the irradiation) and were used to constrain its radial
velocity semi-amplitude using Equation~\ref{eqn:kcorr}. These lines (in
conjunction with the white dwarf's features) can also be used to measure the
gravitational redshift of the white dwarf and hence constrain the physical
parameters of the stars.

The mass and radius of the white dwarf were determined to be
$0.416\pm0.036${\MSUN} and $0.0252\pm0.0017${\RSUN}. Despite the
relatively large uncertainties on these measurements, the radius is
consistent with He core models, but substantially oversized for C/O core models
(by 25 per cent). The hydrogen layer mass cannot be constrained from our
measurements. 

\subsection{CSS\,21357}

CSS\,21357 (SDSS\,J134841.61+183410.5 in SIMBAD) was discovered by
\citet{drake10} to be an eclipsing binary containing a 16,000\,K white dwarf
and an M3 main-sequence star in a 6.0\,h binary. Clean features from both stars
are visible in the X-shooter spectra and the rotational broadening of the M
star was used to constrain the system inclination.

The white dwarf's mass and radius are $0.658\pm0.010${\MSUN} and
$0.0122\pm0.0005${\RSUN}. The radius is consistent with C/O core models,
though the uncertainty is too large to constrain the hydrogen layer mass.

\subsection{CSS\,40190}

CSS\,40190 (SDSS\,J083845.86+191416.5 in SIMBAD) is an eclipsing system
consisting of a 14,900\,K white dwarf and an M5 main-sequence star in a 3.1\,h
period discovered by \citet{drake10}. While the white dwarf dominates the
optical spectrum, features from the main-sequence star are still visible and
the star is not strongly irradiated, therefore reliable velocities could be
measured from both stars. Gravitational redshift and rotational broadening
measurements were also made, with the broadening measurement used to constrain
the inclination.

The resultant mass and radius measurements for the white dwarf were
$0.482\pm0.008${\MSUN} and $0.0158\pm0.0004${\RSUN}, placing it in a
region where both C/O and He core white dwarfs may reside. The measured radius
is marginally more consistent with a C/O core white dwarf, but a He core cannot
be ruled out in this case, as shown in Figure~\ref{fig:mr_conts}.

\subsection{RR\,Cae}

\begin{figure}
  \begin{center}
    \includegraphics[width=\columnwidth]{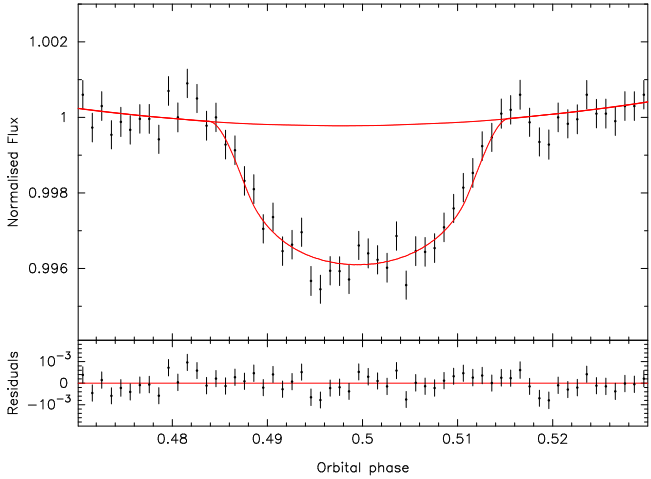}
    \caption{ULTRACAM $i'$ band light curve of the secondary eclipse of
      RR\,Cae (i.e. the transit of the white dwarf across the face of the
      main-sequence star). The best fit model is shown in red as well as a
      model in which the secondary eclipse has been turned off, demonstrating
      its shallow depth and also the small amount of ellipsoidal modulation
      present (which causes the curve in the out-of-eclipse data).} 
  \label{fig:rrcae2e}
  \end{center}
\end{figure}

RR\,Cae is a bright ($V=14.4$) close-by (20pc, \citealt{subasavage09}) cool
7,540\,K DAZ white dwarf in a 7.3\,h binary with an M4 main-sequence star,
discovered to be eclipsing by \citet{krzeminski84}. We did not
observe RR\,Cae spectroscopically with X-shooter because archival high quality
UVES data already exist and were thoroughly analysed by \citet{ribeiro13}. We
use their spectroscopic constraints (listed in Table~\ref{tab:specmeas}) in
addition to our own light curve data to constrain the system parameters.

RR\,Cae is the only one of our systems in which the secondary eclipse is
visible (though only in the $i'$ band) and can be used to constrain the
inclination and radii of both stars; note that several previously
published systems also show secondary eclipses: NN\,Ser \citep{parsons10} and
SDSS\,J0857+0342 \citep{parsons12sp}. In Figure~\ref{fig:rrcae2e} we show the
secondary eclipse light curve, which is a combination of six separate
observations. Additionally, the main-sequence star in RR\,Cae is particularly
active and often flares, hence two observations had to be excluded due to
flares interfering with the eclipse. Nevertheless, the secondary eclipse is
well detected (despite its shallow $\sim$0.4 per cent depth) and, in
combination with the primary eclipse and spectroscopic data, places firm
constraints on the mass and radius of the white dwarf of
$0.448\pm0.002${\MSUN} and $0.01568\pm0.00009${\RSUN}. The measured radius
is consistent with the predictions of He core white dwarf models
with a thick surface hydrogen layer, but is slightly oversized for a C/O core
white dwarf.

\subsection{SDSS\,J0024+1745}

SDSS\,J0024+1745 (SDSS\,J002412.87+174531.4 in SIMBAD) was discovered as an
eclipsing binary by \citet{parsons15}. It shows the shallow eclipse of a cool
8,300\,K white dwarf by an M2.5 main-sequence star in a 4.8\,h orbit. The
main-sequence star dominates the optical spectrum, but white dwarf features
are visible in the X-shooter spectrum, specifically the narrow \Ion{Ca}{ii} H
and K lines, which yielded high precision velocity measurements. The
rotational broadening of the M star was used to constrain the orbital
inclination and physical parameters of the stars.

We determined the mass and radius of the white dwarf to be
$0.534\pm0.009${\MSUN} and $0.0140\pm0.0007${\RSUN}, consistent
with C/O core models, though not precise enough to constrain the hydrogen layer
mass.  

\subsection{SDSS\,J0106$-$0014}

SDSS\,J0106$-$0014 (SDSS\,J010623.01$-$001456.3 in SIMBAD) was discovered to
be an eclipsing binary by \citet{kleinman04}, who noted the discrepancy
between its SDSS photometric and spectroscopic magnitudes, indicating that the
system was at least partially in eclipse during the photometric
observations. It consists of a 14,000\,K white dwarf and a very low-mass M
dwarf companion in a short 2.0\,h period.

Our spectroscopic data revealed absorption lines from the main-sequence star
that allowed us to trace its motion, but were too weak to give reliable
rotational broadening measurements. Therefore, we used the gravitational
redshift of the white dwarf to help constrain the binary inclination. The mass
and radius of the white dwarf were determined to be $0.441\pm0.014${\MSUN}
and $0.0175\pm0.0008$\RSUN, consistent with He core models (although not
precise enough to constrain the hydrogen layer mass) and slightly oversized
for C/O core models, although overlapping with thick hydrogen layer C/O
  core models at the 2$\sigma$ level.

\subsection{SDSS\,J0110+1326}

\begin{figure}
  \begin{center}
    \includegraphics[width=\columnwidth]{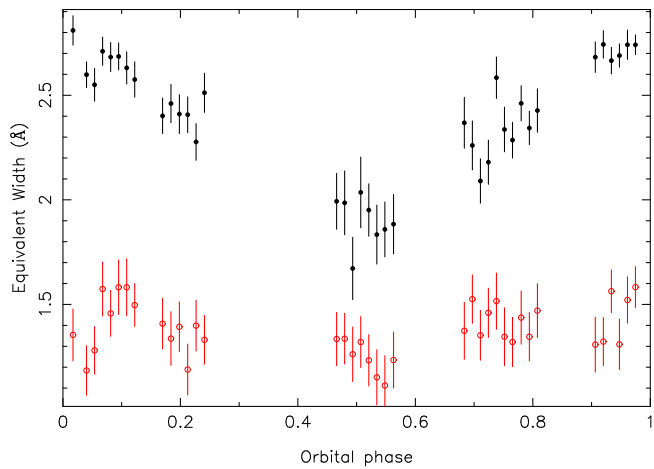}
    \caption{Measured equivalent width of the \Ion{Na}{i} 8183/94{\AA}
      absorption doublet (filled black points) and the \Ion{K}{i} 12500{\AA}
      absorption line (open red points) originating from the heated
      main-sequence star in SDSS\,J0110+1326. The sodium doublet clearly
      becomes weaker around phase 0.5, when the heated hemisphere of the star
      is visible, implying that these lines are affected by the irradiation
      from the white dwarf and hence do not reliably track the centre-of-mass
      of the main-sequence star. The potassium line shows no such variations
      in strength and therefore can be used to properly trace the motion of
      the star.}
  \label{fig:sdss0110ew}
  \end{center}
\end{figure}

SDSS\,J0110+1326 (SDSS\,J011009.08+132616.7 in SIMBAD) is an eclipsing binary
containing a relatively hot 24,500\,K white dwarf and an M4.5 main-sequence
star with a period of 8.0\,h discovered by \citet{pyrzas09}. The main-sequence
star is moderately irradiated and therefore many of its absorption lines are
actually slightly filled in at different orbital phases thus rendering them
unreliable for tracking the centre-of-mass of the star. This was previously
seen by \citet{pyrzas09} who noted the equivalent width of the main-sequence
star's lines were highly phase dependent. However, our X-shooter data cover a
much wider wavelength range than the \citet{pyrzas09} spectroscopy. Crucially,
our data cover the near-infrared spectral range where several strong potassium
lines are located. The \Ion{K}{i} 12500{\AA} absorption line from the
main-sequence star shows no variation in its equivalent width over the orbital
period and therefore appears to be unaffected by the irradiating flux from the
white dwarf (see Figure~\ref{fig:sdss0110ew}), allowing us to reliably trace
its centre-of-mass as well as measure its rotational broadening. Note that the
radial velocity measurement from the sodium doublet lines was
$194.2\pm2.7$\kms, slightly larger than the $190.9\pm1.2${\kms} measured
from the potassium line, demonstrating the distorting effect of irradiation.

The white dwarf's mass and radius are $0.466\pm0.009${\MSUN} and
$0.0184\pm0.0004${\RSUN}, consistent with He core models but slightly
oversized for C/O core models. The radius is also marginally more consistent
with a thick hydrogen layer mass.

\subsection{SDSS\,J0314+0206}

SDSS\,J0314+0206 (WD\,0312+019 in SIMBAD) was discovered by \citet{drake14}
to be a deeply eclipsing binary containing a hot 46,800\,K white dwarf with an
M3 main-sequence star companion in a 7.3\,h orbit. Despite the high temperature
of the white dwarf, the main-sequence star's features are still visible in the
X-shooter spectra, thanks to its relatively high mass. However, it is
substantially irradiated by the white dwarf and there are numerous emission
lines throughout the spectra and many of the absorption lines show variations
in their strength at different orbital phases. Nevertheless, like
SDSS\,J0110+1326, the \Ion{K}{i} 12500{\AA} absorption line in the
near-infrared shows no variations in strength throughout the orbit and
therefore can be used to reliably track the motion of the main-sequence
star. However, the absorption line is quite weak, since it is heavily
  diluted by the white dwarf's flux, to the extent that the rotational
  broadening measurement suffered from substantial uncertainty and hence would
  not strongly  constrain the inclination, compared to other methods.
Therefore, we relied on the gravitational redshift measurement to constrain
the system parameters. As well as the standard Balmer absorption lines, the
white dwarf spectrum also contains \Ion{He}{ii} 4686{\AA} and \Ion{He}{i}
5876{\AA} absorption lines (hence this is a DAO white dwarf), the latter of
which is relatively narrow allowing precise velocity measurements.

The white dwarf has a mass and radius of $0.596\pm0.009${\MSUN} and
$0.0159\pm0.0002${\RSUN} and has a radius consistent with C/O core models
with a thick hydrogen envelope, as demonstrated in Figure~\ref{fig:mr_conts}.

\subsection{SDSS\,J1021+1744}

SDSS\,J1021+1744 (SDSS\,J102102.25+174439.9 in SIMBAD) is a 10,600\,K white
dwarf in a 3.4\,h binary with an M3 main-sequence companion discovered to
eclipse by \citet{parsons13}. Away from the eclipse, the light curve shows
deep, sharp dips caused by prominence like structures originating from the M
star that pass in front of the white dwarf \citep{irawati16}. We removed all
the dips from each ULTRACAM light curve individually (by simply excluding any
points obviously below the mean out-of-eclipse flux, plus a small buffer of 20
seconds either side of a dip), before combining them to make the final eclipse
light curve. The three ULTRACAM observations of SDSS\,J1021+1744 were 
separated by more than four months and, as noted in \citet{irawati16}, the
dips vary in strength and position over long timescales, therefore, between
the three observations we were able to construct a clean eclipse light curve.

While the M star's features are easily visible in the X-shooter spectra, the
white dwarf's features are heavily diluted by the M star. However, the white
dwarf accretes material from the M star's wind resulting in emission lines
from the white dwarf in the Balmer series allowing us to track its
motion (these emission features have been seen in other detached white dwarf
plus main-sequence star binaries, e.g. \citealt{tappert11,parsons12uc}). We
used the rotational broadening measurement to constrain the system parameters.

The white dwarf has a mass and radius of $0.534\pm0.004${\MSUN} and
$0.0140\pm0.0003${\RSUN}, consistent with C/O core models. The radius is also
marginally more consistent with thick hydrogen layer models. Interestingly,
our measurements place the white dwarf in SDSS\,J1021+1744 within the ZZ\,Ceti
instability strip, although the uncertainty in temperature is large enough
that it may still fall outside the strip. However, we see no evidence in our
data for pulsations nor were there any obvious signs of pulsations in the
long term monitoring data in \citet{irawati16}. 

\subsection{SDSS\,J1028+0931}

SDSS\,J1028+0931 (SDSS\,J102857.78+093129.8 in SIMBAD) is a binary containing
a 12,200\,K white dwarf and an M2.5 main-sequence star in a 5.6\,h orbit
discovered to eclipse by \citet{parsons13}. The main-sequence star dominates
the optical flux and hence the eclipse is shallow. However, white dwarf
features are visible in the X-shooter spectra and, like SDSS\,J1021+1744,
accretion of wind material drives emission lines in the Balmer series that
were used to track the white dwarf's motion. We used the rotational broadening
measurement to constrain the system parameters.

We determined the mass and radius of the white dwarf to be
$0.415\pm0.004${\MSUN} and $0.0177\pm0.0002${\RSUN}, consistent with He
core models but oversized for C/O models. The radius is also more consistent
with thick hydrogen layer models than those with thinner layers. These
parameters place the white dwarf close to the blue edge of the ZZ\,Ceti
instability strip, but our (limited) photometry does not show any signs of
pulsations.

\subsection{SDSS\,J1123$-$1155}

Discovered to be an eclipsing binary by \citet{parsons15}, SDSS\,J1123$-$1155
(SDSS\,J112308.40$-$115559.3 in SIMBAD) consists of a 10,200K white dwarf in an
18.5\,h orbit with a M3.5 main-sequence star, making it the longest period
binary in our sample. Features from both stars are visible in the spectrum
including narrow metal absorption lines in the white dwarf's spectrum caused
by accreting wind material from the main-sequence star. However, due to the
much longer period of this system, compared to the others in our sample, the
rotational broadening of the M star is relatively small and in fact cannot be
measured with the resolution of the X-shooter data. Therefore, we rely on the
gravitational redshift measurement to determine the inclination and hence
physical parameters of the stars.

The white dwarf in SDSS\,J1123$-$1155 was found to have a mass and radius of
$0.605\pm0.008${\MSUN} and $0.0128\pm0.0004${\RSUN}, consistent
with C/O core models. However, the measurements are not precise enough to
constrain the hydrogen layer mass. 

\subsection{SDSS\,J1307+2156}

SDSS\,J1307+2156 (SDSS\,J130733.49+215636.7 in SIMBAD) contains a cool
8,500\,K white dwarf in a 5.2\,h binary with an M4 main-sequence star and was
discovered to be eclipsing by \citet{parsons13}. Originally listed as a DC
white dwarf from its SDSS spectrum \citep{rebassa12}, our higher resolution
X-shooter data revealed narrow Balmer absorption lines as well as a large
number of metal lines caused by wind accretion. The M star dominates the flux
redward of H$\beta$, hence we use the measurement of its rotational
broadening to constrain the inclination and physical parameters of the
binary.

We determined the mass and radius of the white dwarf to be
$0.610\pm0.003${\MSUN} and $0.0121\pm0.0006${\RSUN}, consistent with C/O
core models, though not precise enough to constrain the hydrogen layer mass.

\subsection{SDSS\,J1329+1230}

SDSS\,J1329+1230 (SDSS\,J132925.21+123025.4 in SIMBAD) was found to be an
eclipsing binary by \citet{drake10}. It contains a 12,500\,K white dwarf with
a low-mass M8 companion with a short 1.9\,h period. The white dwarf completely
dominates the optical spectrum and narrow calcium and magnesium absorption
lines are visible in its spectrum due to it accreting some material from the
wind of its companion. No absorption features are visible from the M star,
though several emission lines are seen, arising from the heating effect
from the white dwarf. We use these and Equation~\ref{eqn:kcorr} to determine
its radial velocity and use the gravitational redshift of the white dwarf to
constrain the inclination and hence physical parameters of the stars.

The mass and radius of the white dwarf were found to be
$0.392\pm0.023${\MSUN} and $0.0180\pm0.0005${\RSUN}, making this the
lowest mass white dwarf in our sample. The radius is consistent with He core
models, but slightly too large for C/O core models (4 per cent oversized, see
Figure~\ref{fig:mr_conts}). The precision of the measurements is insufficient
to constrain the hydrogen layer mass.

\begin{table*}
 \centering
  \caption{White dwarf mass-radius measurements obtained from eclipsing
    binaries. $R/R_\mathrm{C/O}$ is the ratio of the measured radius to the
    theoretical radius of a C/O core white dwarf with a thick
    ($M_\mathrm{H}/M_\mathrm{WD}=10^{-4}$) hydrogen envelope (taken from
    \citealt{fontaine01} or \citealt{benvenuto99} for
    $T_\mathrm{eff}>40,000$\,K), $R/R_\mathrm{He}$ is the ratio of the
    measured radius to the theoretical radius of a He core white dwarf with a
    thick ($M_\mathrm{H}/M_\mathrm{WD}=10^{-4}$) hydrogen envelope
    \citep{panei07}, for white dwarfs with masses less than
    0.5\MSUN. References: (1) This paper, (2) \citet{bours14}, (3)
    \citet{parsons12}, (4) \citet{parsons10}, (5) \citet{parsons16}, (6)
    \citet{parsons12uc}, (7) \citet{parsons12sp}, (8) \citet{pyrzas12}, (9)
    \citet{obrien01}.} 
  \label{tab:wdparas}
  \begin{tabular}{@{}lcccccccc@{}}
  \hline
  Object             & $g$ mag & P$_\mathrm{orb}$ (h) & $T_\mathrm{eff}$ (K) & Mass (\MSUN)     & Radius (\RSUN)       & $R/R_\mathrm{C/O}$ & $R/R_\mathrm{He}$ & Ref \\
  \hline
  CSS\,080502        & 17.08   & 3.587  & $17838\pm482$      & $0.4756\pm0.0036$ & $0.01749\pm0.00028$ & $1.06\pm0.01$    & $0.98\pm0.01$   & 1   \\
  CSS\,09704         & 18.41   & 3.756  & $29969\pm679$      & $0.4164\pm0.0356$ & $0.02521\pm0.00170$ & $1.25\pm0.08$    & $1.05\pm0.08$   & 1   \\
  CSS\,21357         & 17.29   & 5.962  & $15909\pm285$      & $0.6579\pm0.0097$ & $0.01221\pm0.00046$ & $0.98\pm0.03$    & -               & 1   \\
  CSS\,40190         & 18.16   & 3.123  & $14901\pm731$      & $0.4817\pm0.0077$ & $0.01578\pm0.00039$ & $1.02\pm0.03$    & $0.95\pm0.03$   & 1   \\
  CSS\,41177A        & 17.27   & 2.784  & $22497\pm60$       & $0.3780\pm0.0230$ & $0.02224\pm0.00041$ & $1.09\pm0.02$    & $0.98\pm0.02$   & 2   \\
  CSS\,41177B        & 17.27   & 2.784  & $11864\pm281$      & $0.3160\pm0.0110$ & $0.02066\pm0.00042$ & $1.05\pm0.02$    & $0.97\pm0.02$   & 2   \\
  GK\,Vir            & 16.81   & 8.264  & $50000\pm673$      & $0.5618\pm0.0142$ & $0.01700\pm0.00030$ & $0.96\pm0.01$    & -               & 3   \\
  NN\,Ser            & 16.43   & 3.122  & $63000\pm3000$     & $0.5354\pm0.0117$ & $0.02080\pm0.00020$ & $1.00\pm0.01$    & -               & 4   \\
  QS\,Vir            & 14.66   & 3.618  & $14220\pm350$      & $0.7816\pm0.0130$ & $0.01068\pm0.00007$ & $1.00\pm0.01$    & -               & 5   \\
  RR\,Cae            & 14.57   & 7.289  & $7540\pm175$       & $0.4475\pm0.0015$ & $0.01568\pm0.00009$ & $1.03\pm0.01$    & $1.00\pm0.01$   & 1   \\
  SDSS\,J0024$+$1745 & 18.71   & 4.801  & $8272\pm580$       & $0.5340\pm0.0090$ & $0.01398\pm0.00070$ & $1.01\pm0.05$    & -               & 1   \\
  SDSS\,J0106$-$0014 & 18.14   & 2.040  & $13957\pm531$      & $0.4406\pm0.0144$ & $0.01747\pm0.00083$ & $1.05\pm0.01$    & $1.00\pm0.01$   & 1   \\
  SDSS\,J0110$+$1326 & 16.53   & 7.984  & $24569\pm385$      & $0.4656\pm0.0091$ & $0.01840\pm0.00036$ & $1.07\pm0.02$    & $0.99\pm0.02$   & 1   \\
  SDSS\,J0138$-$0016 & 18.84   & 1.746  & $3570\pm100$       & $0.5290\pm0.0100$ & $0.01310\pm0.00030$ & $0.99\pm0.02$    & -               & 6   \\
  SDSS\,J0314$+$0206 & 16.95   & 7.327  & $46783\pm7706$     & $0.5964\pm0.0088$ & $0.01594\pm0.00022$ & $1.01\pm0.01$    & -               & 1   \\
  SDSS\,J0857$+$0342 & 17.95   & 1.562  & $37400\pm400$      & $0.5140\pm0.0490$ & $0.02470\pm0.00080$ & $1.43\pm0.06$    & $1.08\pm0.06$   & 7   \\
  SDSS\,J1021$+$1744 & 19.51   & 3.369  & $10644\pm1721$     & $0.5338\pm0.0038$ & $0.01401\pm0.00032$ & $1.00\pm0.02$    & -               & 1   \\
  SDSS\,J1028$+$0931 & 16.40   & 5.641  & $12221\pm765$      & $0.4146\pm0.0036$ & $0.01768\pm0.00020$ & $1.07\pm0.01$    & $1.02\pm0.01$   & 1   \\
  SDSS\,J1123$-$1155 & 17.99   & 18.459 & $10210\pm87$       & $0.6050\pm0.0079$ & $0.01278\pm0.00037$ & $0.99\pm0.02$    & -               & 1   \\
  SDSS\,J1210$+$3347 & 16.94   & 2.988  & $6000\pm200$       & $0.4150\pm0.0100$ & $0.01590\pm0.00050$ & $1.02\pm0.03$    & $0.99\pm0.03$   & 8   \\
  SDSS\,J1212$-$0123 & 16.77   & 8.061  & $17707\pm35$       & $0.4393\pm0.0022$ & $0.01680\pm0.00030$ & $1.00\pm0.01$    & $0.94\pm0.01$   & 3   \\
  SDSS\,J1307$+$2156 & 18.25   & 5.192  & $8500\pm500$       & $0.6098\pm0.0031$ & $0.01207\pm0.00061$ & $0.96\pm0.04$    & -               & 1   \\
  SDSS\,J1329$+$1230 & 17.26   & 1.943  & $12491\pm312$      & $0.3916\pm0.0234$ & $0.01800\pm0.00052$ & $1.04\pm0.02$    & $0.99\pm0.02$   & 1   \\
  SDSS\,J2235$+$1428 & 18.59   & 3.467  & $20837\pm773$      & $0.3977\pm0.0220$ & $0.01975\pm0.00050$ & $1.08\pm0.02$    & $0.98\pm0.02$   & 1   \\
  V471\,Tau          & 10.04   & 12.508 & $34500\pm1000$     & $0.8400\pm0.0500$ & $0.01070\pm0.00070$ & $1.06\pm0.06$    & -               & 9   \\
  WD\,1333+005       & 17.41   & 2.927  & $7740\pm73$        & $0.4356\pm0.0016$ & $0.01570\pm0.00036$ & $1.02\pm0.02$    & $0.98\pm0.02$   & 1   \\
  \hline
\end{tabular}
\end{table*}

\subsection{SDSS\,J2235+1428}

SDSS\,J2235+1428 (SDSS\,J223530.62+142855.1 in SIMBAD) contains a 20,800\,K
white dwarf in a 3.5\,h binary with an M5 main-sequence star and was discovered
to be eclipsing by \citet{parsons13}. The white dwarf dominates the optical
spectrum, although features from the M star are visible redward of
H$\alpha$. The M star is moderately heated and several emission lines are
visible throughout the spectrum. However, the sodium 8200{\AA} absorption
doublet appears to be unaffected by the heating, since the equivalent widths
of the two lines show no variation with orbital phase. Therefore, we use these
to track the main-sequence star's motion and measure its rotational
broadening in order to constrain the inclination.

The white dwarf has a mass and radius of $0.398\pm0.022${\MSUN} and
$0.0198\pm0.0005${\RSUN}, consistent with He core models, but substantially
oversized for a C/O core white dwarf (8 per cent too large). The radius
measurement is slightly more consistent with a thick hydrogen layer, but a
thinner layer cannot be excluded.

\subsection{WD\,1333+005}

WD\,1333+005 was found to be an eclipsing binary by \citet{drake10}. It
contains a cool 7,700\,K white dwarf and M5 main-sequence star with a 2.9\,h
period. The white dwarf's spectrum contains a huge number of narrow metal
absorption lines due to accreting material from the wind of the main-sequence
star. The M star dominates the flux beyond H$\beta$ and therefore we use the
measurement of its rotational broadening to constrain the system parameters.

The mass and radius of the white dwarf are $0.436\pm0.002${\MSUN} and
$0.0157\pm0.0004${\RSUN}. Despite the high precision of our measurements,
the radius is in agreement with both He and C/O core models (since the difference
between the models is small at lower temperatures), although the measured
radius is inconsistent with thin hydrogen layer C/O core models.

\section{Discussion}

\begin{figure*}
  \begin{center}
    \includegraphics[width=0.97\columnwidth]{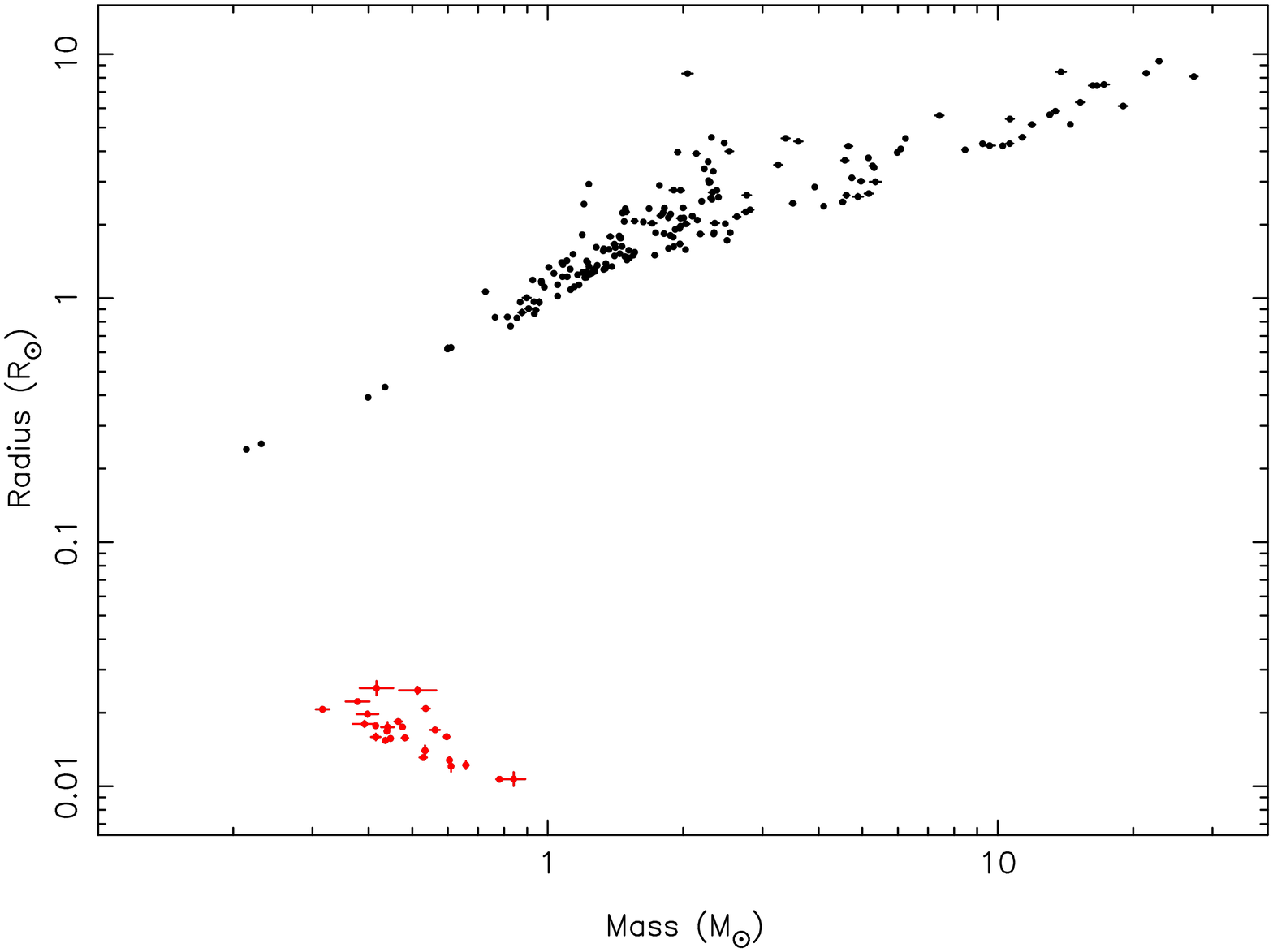}
    \hspace{5mm}
    \includegraphics[width=0.97\columnwidth]{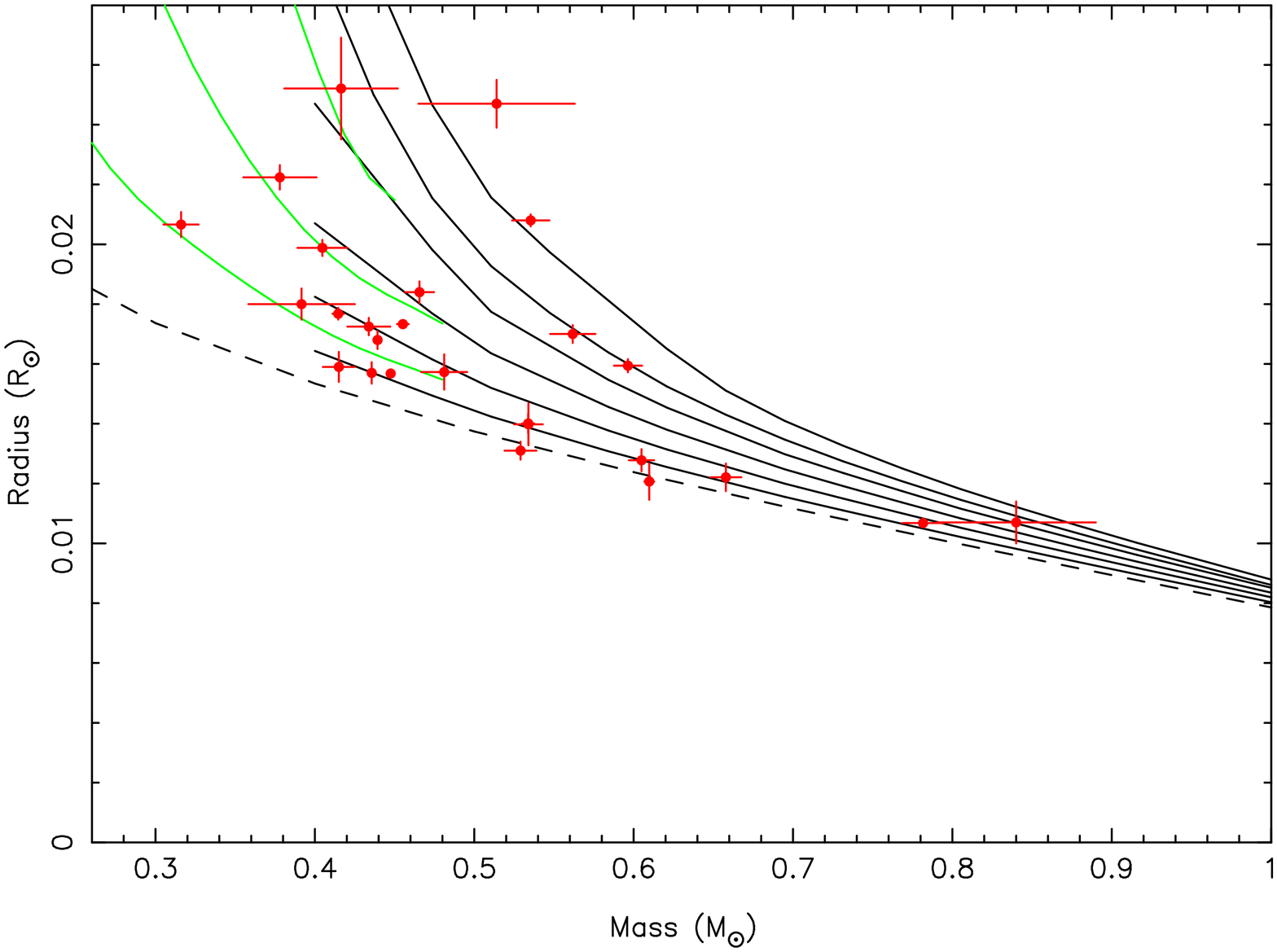}
    \caption{{\it Left:} stars and white dwarfs with precise model-independent
      mass and radius measurements. The white dwarfs are shown in red and are
      from Table~\ref{tab:wdparas}, the main-sequence stars are taken from
      \citet{torres10}. This figure demonstrates not only the inverse
      mass-radius relationship for white dwarfs, but also the large mass
      losses involved in stellar evolution. {\it Right:} The white dwarf
      mass-radius relationship. The solid black lines are theoretical C/O core
      models with thick hydrogen layers with temperatures from 60,000\,K (top)
      to 10,000\,K (bottom) in steps of 10,000\,K (taken from
      \citealt{fontaine01} or \citealt{benvenuto99} for
      $T_\mathrm{eff}>40,000$\,K). The green lines are He core models with 
      thick hydrogen layers with temperatures from 30,000\,K (top) to
      10,000\,K (bottom) in steps of 10,000\,K (taken from
      \citealt{panei07}). The dashed black line is the zero-temperature
      relation from \citet{verbunt88}.}
  \label{fig:WDMR}
  \end{center}
\end{figure*}

We summarise all our measurements in Table~\ref{tab:wdparas} along with other,
previously published precise white dwarf mass-radius measurements from close
binaries. We use the term ``precise'' to mean any mass and radius
  measurements that have been made independent of any theoretical mass-radius
  relationships or spectral fits and with uncertainties less than 10 per
  cent.

\begin{figure}
  \begin{center}
    \includegraphics[width=\columnwidth]{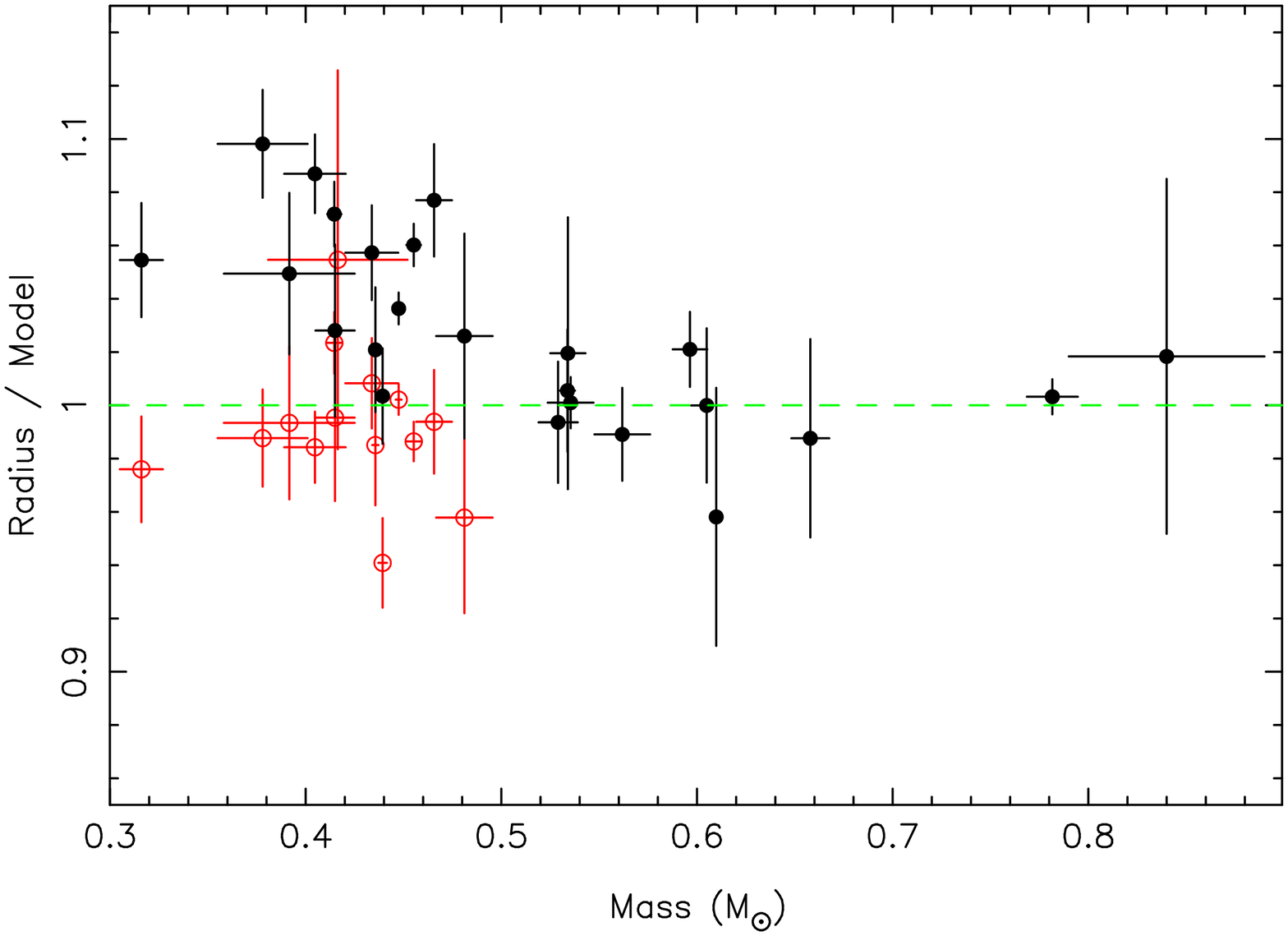}
    \caption{Ratio of the measured white dwarf radii to theoretical
      predictions as a function of mass. Filled black points are assuming a C/O
      core white dwarf with a thick hydrogen layer (taken from
      \citealt{fontaine01} or \citealt{benvenuto99} for
      $T_\mathrm{eff}>40,000$\,K), while open red points are assuming a He
      core with a thick hydrogen layer (for the same white dwarfs, with masses
      below 0.5{\MSUN}, taken from \citealt{panei07}). Virtually every radius
      measurement below 0.5{\MSUN} is more consistent with He core rather than
      C/O core models, the only exception being SDSS\,J1212$-$0123
      \citep{parsons12}.}
  \label{fig:oversize_m}
  \end{center}
\end{figure}

The white dwarfs in Table~\ref{tab:wdparas} span a wide range of masses and
temperatures, as is best seen in Figure~\ref{fig:WDMR}, although there is a
clear lack of massive ($>$1\MSUN) white dwarfs, which is unsurprising given the
evolutionary history of these binaries. Note that these are currently
the only white dwarfs with independent mass and radius measurements, since
white dwarfs in astrometric binaries or common proper motion systems still
rely on spectral fits to determine their emergent flux and hence radii. While
Figure~\ref{fig:WDMR} nicely demonstrates the inverse mass-radius relationship
for white dwarfs (i.e. that more massive white dwarfs are generally smaller
than less massive ones, with some dependence upon temperature), it is not
straightforward to assess the agreement with evolutionary models due to the
different temperatures and core compositions of all these white
dwarfs. Therefore, in Figure~\ref{fig:oversize_m} we plot the ratio of the
measured radius to the theoretical radius (taking into account the uncertainty
in temperature as well), assuming thick surface hydrogen layers. For white
dwarfs with masses less than 0.5{\MSUN} we plot this ratio relative to C/O core
models (in black) and He core models (in red). Both Figure~\ref{fig:mr_conts}
and Figure~\ref{fig:oversize_m} show that white dwarfs with masses below 
0.5{\MSUN} have radii that are more consistent with He core models and only one
object, SDSS\,J1212$-$0123, has a radius more consistent with a C/O core
\citep{parsons12}. It is perhaps unsurprising that these low mass white dwarfs
likely have He cores since, although it is possible to create C/O core white
dwarfs with masses less than 0.5{\MSUN}, it requires substantial mass loss
along the red giant branch \citep{prada09,willems04,han00}. For this to occur
in a binary system requires that the initial mass ratio was large and
therefore the orbit should increase during the mass loss and hence we would
not observe such systems as close binaries. Nevertheless, this is the first
direct observational evidence of a change in core composition in this mass
range. With a sufficiently large sample of mass-radius measurements in this
range one could even pinpoint exactly where this core change occurs
(although this also depends upon the metallicity and so will not be the same
for every object).

\begin{figure}
  \begin{center}
    \includegraphics[width=\columnwidth]{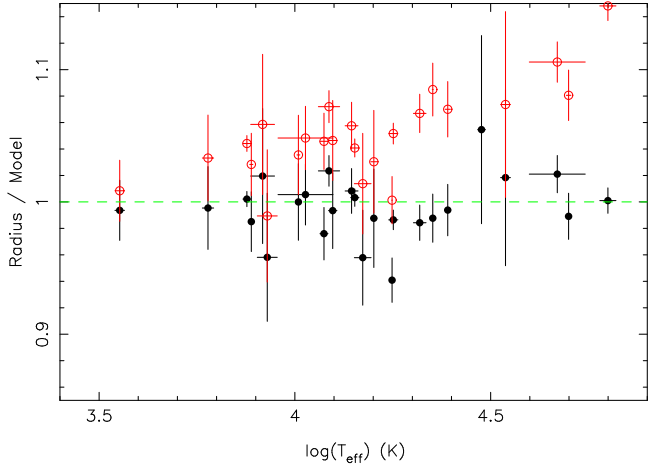}
    \caption{Ratio of the measured white dwarf radii to theoretical
      predictions as a function of effective temperature. Filled black points
      assume a thick surface hydrogen layer, while open red points assume a
      thin layer. The discrepancy is largest for hotter white dwarfs, with the
      majority of measurements favoring a thick layer.}
  \label{fig:oversize_t}
  \end{center}
\end{figure}

In addition to mass, temperature and core composition, the radius of a white
dwarf is also dependent upon the surface hydrogen layer. Studies of pulsating
ZZ\,Ceti type white dwarfs have found a large spread in the hydrogen layer
mass of white dwarfs, from as low as $M_\mathrm{H}/M_\mathrm{WD}=10^{-10}$ upto
$M_\mathrm{H}/M_\mathrm{WD}=10^{-4}$ \citep{castanheira09,romero12}, with a
mean value of $M_\mathrm{H}/M_\mathrm{WD}=2.71\times10^{-5}$. While our
measurements cannot constrain the layer mass as precisely as pulsational
studies, we can still potentially differentiate between thick and thin
layers. In Figure~\ref{fig:oversize_t} we plot the ratio of the measured white
dwarf radius to the theoretical radius (assuming He cores for masses below
0.5\MSUN) for thick hydrogen layers ($M_\mathrm{H}/M_\mathrm{WD}=10^{-4}$,
black points) and thin hydrogen layers ($M_\mathrm{H}/M_\mathrm{WD}=10^{-10}$, red
points). For objects with precise enough measurements we find that models with
thick hydrogen layers reproduce the observed radii far better than models with
thinner layers (that are often around 10 per cent too small). This effect is
largest in hotter white dwarfs, where the layer mass has a much larger impact
on the overall radius of the white dwarf.

\begin{figure}
  \begin{center}
    \includegraphics[width=\columnwidth]{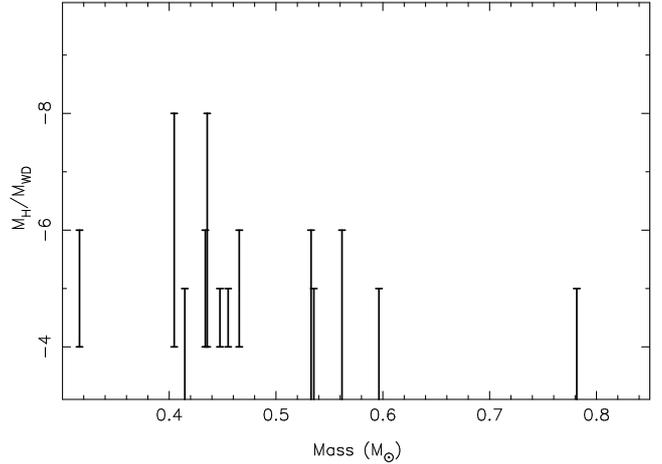}
    \caption{Range of surface hydrogen layer masses consistent with our
      measured radii. We have excluded objects where we have minimal or no
      constraints.} 
  \label{fig:layermass}
  \end{center}
\end{figure}

We also compared our measured radii to the models of \citet{benvenuto99} (C/O
core) and \citet{panei07} (He core) over a range of different hydrogen
layer masses (note that the \citealt{fontaine01} models used in
Figure~\ref{fig:oversize_t} are only available with thick,
$M_\mathrm{H}/M_\mathrm{WD}=10^{-4}$, or thin,
$M_\mathrm{H}/M_\mathrm{WD}=10^{-10}$ layers). Model radii within one sigma of
the measured value were deemed consistent (taking into account the
uncertainties in the mass and temperature of each white dwarf) and are shown in
Figure~\ref{fig:layermass}. We find that, for white dwarfs with precise enough
measurements, their radii are consistent with having surface hydrogen layers
with masses between $10^{-5} \ge M_\mathrm{H}/M_\mathrm{WD} \ge 10^{-4}$, with
some spread to larger and smaller values. This is consistent with the
canonical value of $10^{-4}$ from evolutionary computations
\citep[e.g.][]{althaus10}. Since all these white dwarfs have interacted with
their main-sequence companions in the past (during the giant phase, through a
common envelope event) it might be expected that they posses thinner hydrogen
envelopes than isolated white dwarfs, although the small amount of material
accreted from the wind of their companions ($\sim$$10^{-15}\,
\mathrm{M}_{\odot}\, \mathrm{yr}^{-1}$ \citealt{pyrzas12}) ensures that there 
are no DB white dwarfs in close binaries with main-sequence stars
\citep{parsons13}. However, the fact that our measurements are consistent
with the canonical value implies that the hydrogen envelopes are not strongly
affected by common envelope evolution. On the other hand, the hydrogen layer
mass of the ZZ\,Ceti white dwarf in the close binary SDSS\,J113655.17+040952.6
was found to be $M_\mathrm{H}/M_\mathrm{WD} \approx 10^{-4.9}$
\citep{hermes15}, at the lower end of the range we find and potentially
indicative of some small mass loss. A more robust test of whether or not the
common envelope phase has an effect on the layer mass will require the
identification of more ZZ\,Ceti white dwarfs in close binaries.  

\subsection{The accuracy of white dwarf spectroscopic fits}

For the vast majority of single white dwarfs, spectroscopic fits to
the Balmer lines are currently the only method to estimate their
masses (using the surface gravity determined from the fits and a
mass-radius relation), and most of our knowledge of the white dwarf
mass distribution is based on this method
\citep[e.g.][]{bergeron92, finley97, kepler07, tremblay13}. Our sample of
white dwarfs in eclipsing binaries provides an excellent opportunity to test
the precision and robustness of the spectrosopic fits, as we can compare the
surface gravities derived from the X-Shooter spectra with those computed from
the masses and radii that we directly measured from the light curve fits.

\begin{figure}
  \begin{center}
  \includegraphics[width=0.95\columnwidth]{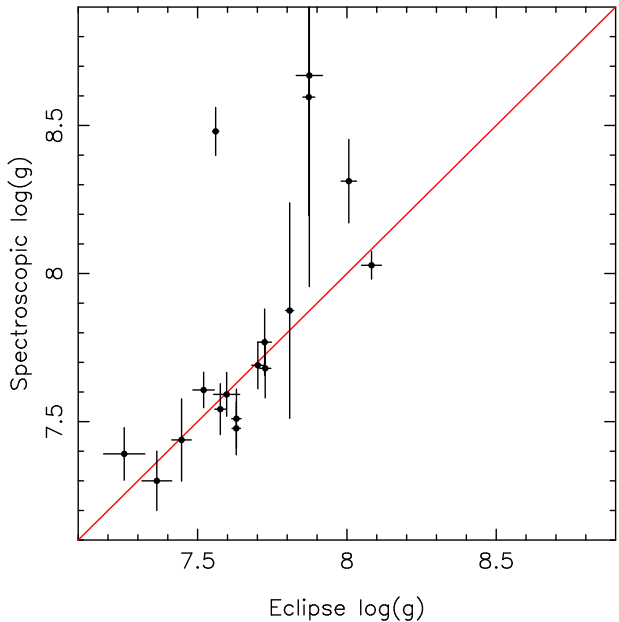}
  \caption{Comparison between the surface gravities determined from the model
    atmosphere fits to the X-Shooter spectroscopy, and those computed from the
    mass and radius values measured from the light curves.}
  \label{fig:complogg}
 \end{center}
\end{figure}

Figure~\ref{fig:complogg} shows the comparison between the surface gravities
of our white dwarfs from the spectrosopic fits and computed from the mass and
radius values measured from the light curves. In general there is excellent
agreement between the spectroscopic surface gravities and the measured
values, however, there are a few outliers and in all these cases the
spectroscopic fit over-predicts the surface gravity. Three of these
objects (SDSS\,J0024+1745, SDSS\,J1021+1744 and SDSS\,J1028+0931) are cool
white dwarfs with early M-type companions that dominate the optical flux over
the white dwarf. Therefore, the white dwarf spectra are of lower quality than
the other objects and may still be contaminated by features from the
main-sequence stars (as noted in Section~\ref{sec:specfit}, the main-sequence
star subtraction is never perfect). This means that the spectroscopic fits are
less reliable, since the Balmer lines may be distorted and hence it is perhaps
unsuprising that these particular objects are outliers.

 The one other outlier is SDSS\,J1123$-$1155, which differs from the other
systems because the white dwarf dominates the optical light and therefore
suffers from less contamination than the other objects. However, in this case
the discrepancy is only at the 2 sigma level (spectroscopic $\log
g=8.31\pm0.14$, eclipse analysis $\log g=8.01\pm0.03$). As the model
atmosphere fit measures the surface gravity from the width and shape of the
Balmer line profiles, additional broadening mechanisms such as additional
perturbers contributing the the Stark broadening (e.g. neutral He), or a
magnetic field, may be the cause for the slightly higher spectroscopically
determined mass.

\section{Conclusions}

We have used a combination of high-speed photometry and phase-resolved
spectroscopy to precisely measure the masses and radii for 16 white dwarfs in
detached eclipsing binaries with low-mass main-sequence star companions. We
combined these results with 10 previously measured binaries to test the white
dwarf mass-radius relationship. We found excellent agreement between our
measured radii and theoretical predictions across a wide range of masses and
temperatures. We also find that, as expected, the radii of white dwarfs with
masses below 0.5{\MSUN} are far more consistent with He core models than those
with C/O cores. Moreover, our most precise measurements allow us to exclude
thin surface hydrogen layers, with thicker values of $10^{-5} \ge
M_\mathrm{H}/M_\mathrm{WD} \ge 10^{-4}$ favoured, implying that surface hydrogen
layers are not strongly depleted by close binary evolution.

\section*{Acknowledgements}

We thank the referee for useful comments and suggestions. SGP acknowledges the
support of the Leverhulme Trust. The research leading to these results has
received funding from the European Research Council under the European Union's
Seventh Framework Programme (FP/2007-2013) / ERC Grant Agreement numbers
340040 (HiPERCAM) and 320964 (WDTracer). ULTRACAM, TRM, VSD, SPL and EB are
supported by the Science and Technology Facilities Council (STFC). MCPB
acknowledges support from the Joint Committee ESO-Government of Chile (grant
2014). ARM acknowledges financial support from MINECO grant
AYA2014-59084-P. Support for this work was provided by NASA through Hubble
Fellowship grant \#HST-HF2-51357.001-A. MRS thanks for support from FONDECYT
(1141269) and Millennium Science Initiative, Chilean ministry of Economy:
Nucleus P10-022-F. This work has made use of data obtained at the Thai
National Observatory on Doi Inthanon, operated by NARIT. The results presented
in this paper are based on observations collected at the European Southern
Observatory under programme IDs 086.D-0161, 086.D-0265 and 192.D-0270. We
thank Alex Gianninas for computing white dwarf limb darkening parameters for
the $KG5$ filter. 

\bibliographystyle{mnras}
\bibliography{wd_mr}

\appendix

\section{Journal of photometric observation}

\begin{table*}
 \centering
  \caption{Journal of photometric observations.}
  \label{tab:photlog}
  \begin{tabular}{@{}lcccccccc@{}}
  \hline
  Date at     &Instrument &Telescope &Filter(s) &Start     &Orbital     &Exposure &Number of &Conditions              \\
  start of run&           &          &          &(UT)      &phase       &time (s) &exposures &(Transparency, seeing)  \\
  \hline
  \multicolumn{2}{l}{\bf CSS\,080502:}\\
  2010-04-21  & ULTRACAM  & NTT      & $u'g'i'$ & 23:48:00 & 0.90--1.60 & 3.0     & 3339     & Good, $\sim$2 arcsec \\
  2010-04-22  & ULTRACAM  & NTT      & $u'g'i'$ & 23:25:47 & 0.48--0.92 & 2.0     & 2734     & Good, $\sim$1.5 arcsec \\
  2010-04-25  & ULTRACAM  & NTT      & $u'g'i'$ & 01:04:33 & 0.33--0.65 & 2.0     & 2066     & Average, $\sim$1.5 arcsec \\
  2010-11-23  & ULTRACAM  & NTT      & $u'g'i'$ & 07:35:09 & 0.79--1.16 & 3.7     & 1269     & Good, $\sim$1.5 arcsec \\
  2010-11-26  & ULTRACAM  & NTT      & $u'g'i'$ & 07:10:17 & 0.75--1.13 & 3.8     & 1269     & Excellent, $<$1 arcsec \\
  2010-11-27  & ULTRACAM  & NTT      & $u'g'i'$ & 06:15:06 & 0.19--0.54 & 3.7     & 1194     & Excellent, $<$1 arcsec \\
  2010-12-15  & ULTRACAM  & NTT      & $u'g'i'$ & 06:52:44 & 0.82--1.29 & 3.7     & 1583     & Good, $\sim$1 arcsec \\
  2010-12-17  & ULTRACAM  & NTT      & $u'g'i'$ & 05:32:20 & 0.83--1.70 & 3.7     & 2964     & Good, $\sim$1 arcsec \\
  2016-11-10  & ULTRACAM  & NTT      & $u'g'r'$ & 06:39:41 & 0.82--1.10 & 3.0     & 1198     & Excellent, $<$1 arcsec \\
  \multicolumn{2}{l}{\bf CSS\,09704:}\\
  2010-11-29  & ULTRACAM  & NTT      & $u'g'i'$ & 00:24:08 & 0.85--1.22 & 6.9     & 740      & Average, $\sim$1.5 arcsec \\
  2011-05-27  & ULTRACAM  & NTT      & $u'g'r'$ & 09:02:36 & 0.88--1.28 & 5.8     & 922      & Good, $\sim$1 arcsec \\
  2011-11-01  & ULTRACAM  & WHT      & $u'g'r'$ & 22:04:39 & 0.91--1.07 & 4.1     & 531      & Excellent, $<$1 arcsec \\
  2012-09-05  & ULTRACAM  & WHT      & $u'g'r'$ & 20:37:30 & 0.91--1.05 & 3.2     & 517      & Good, $\sim$1 arcsec \\
  2012-09-06  & ULTRACAM  & WHT      & $u'g'r'$ & 22:45:51 & 0.87--1.08 & 3.3     & 834      & Average, $\sim$1.5 arcsec \\
  2012-10-09  & ULTRACAM  & WHT      & $u'g'r'$ & 19:43:44 & 0.90--1.11 & 3.1     & 887      & Good, $\sim$1.5 arcsec \\
  2016-11-11  & ULTRACAM  & NTT      & $u'g'r'$ & 01:16:02 & 0.95--1.05 & 4.0     & 318      & Good, $\sim$1 arcsec \\
  \multicolumn{2}{l}{\bf CSS\,21357:}\\
  2014-01-27  & ULTRASPEC & TNT      & $KG5$    & 19:32:23 & 0.92--1.10 & 1.5     & 2484     & Excellent, $\sim$1.5 arcsec \\
  2015-02-28  & ULTRASPEC & TNT      & $KG5$    & 19:26:52 & 0.94--1.09 & 2.0     & 1584     & Good, $\sim$2 arcsec \\
  2015-03-01  & ULTRASPEC & TNT      & $KG5$    & 19:18:13 & 0.94--1.09 & 2.0     & 1670     & Good, $\sim$1.5 arcsec \\
  \multicolumn{2}{l}{\bf CSS\,40190:}\\
  2010-12-13  & ULTRACAM  & NTT      & $u'g'r'$ & 07:41:50 & 0.89--1.14 & 4.8     & 581      & Good, $\sim$2 arcsec \\
  2016-11-11  & ULTRACAM  & NTT      & $u'g'r'$ & 07:32:00 & 0.86--1.12 & 4.0     & 761      & Excellent, $<$1 arcsec \\
  \multicolumn{2}{l}{\bf RR\,Cae:}\\
  2005-11-25  & ULTRACAM  & VLT      & $u'g'i'$ & 00:23:57 & 0.43--0.56 & 0.5     & 6974     & Excellent, $<$1 arcsec \\
  2005-11-27  & ULTRACAM  & VLT      & $u'g'i'$ & 00:02:27 & 0.96--1.06 & 0.5     & 5061     & Excellent, $<$1 arcsec \\
  2005-11-27  & ULTRACAM  & VLT      & $u'g'i'$ & 07:04:42 & 0.93--1.10 & 0.5     & 8617     & Excellent, $<$1 arcsec \\
  2010-11-15  & ULTRACAM  & NTT      & $u'g'i'$ & 03:31:35 & 0.37--0.58 & 3.0     & 1834     & Good, $\sim$1 arcsec \\
  2010-11-21  & ULTRACAM  & NTT      & $u'g'i'$ & 05:34:49 & 0.41--0.85 & 3.0     & 3901     & Good, $\sim$1.5 arcsec \\
  2010-11-22  & ULTRACAM  & NTT      & $u'g'r'$ & 06:07:21 & 0.78--1.09 & 3.0     & 2794     & Good, $\sim$1.5 arcsec \\
  2010-11-26  & ULTRACAM  & NTT      & $u'g'i'$ & 01:43:15 & 0.34--0.56 & 2.5     & 2149     & Excellent, $<$1 arcsec \\
  2010-11-27  & ULTRACAM  & NTT      & $u'g'i'$ & 03:54:33 & 0.94--1.05 & 2.5     & 1216     & Excellent, $<$1 arcsec \\
  2010-11-27  & ULTRACAM  & NTT      & $u'g'i'$ & 07:37:29 & 0.45--0.55 & 2.5     & 1128     & Excellent, $<$1 arcsec \\
  2010-12-02  & ULTRACAM  & NTT      & $u'g'i'$ & 04:09:15 & 0.43--0.57 & 2.5     & 1433     & Good, $\sim$1.5 arcsec \\
  2010-12-10  & ULTRACAM  & NTT      & $u'g'i'$ & 00:43:35 & 0.30--0.62 & 3.0     & 2762     & Good, $\sim$1.5 arcsec \\
  2010-12-15  & ULTRACAM  & NTT      & $u'g'i'$ & 05:26:25 & 0.41--0.59 & 2.8     & 1590     & Good, $\sim$1 arcsec \\
  2016-11-08  & ULTRACAM  & NTT      & $u'g'r'$ & 04:37:46 & 0.98--1.04 & 2.0     & 743      & Excellent, $<$1 arcsec \\
  \multicolumn{2}{l}{\bf SDSS\,J0024+1745:}\\
  2015-09-18  & ULTRACAM  & WHT      & $u'g'r'$ & 02:57:44 & 0.91--1.10 & 0.6     & 5966     & Average, $\sim$2 arcsec \\
  2016-11-08  & ULTRACAM  & NTT      & $u'g'r'$ & 00:13:14 & 0.92--1.07 & 4.0     & 688      & Excellent, $<$1 arcsec \\
  2016-11-09  & ULTRACAM  & NTT      & $u'g'r'$ & 00:31:37 & 0.98--1.09 & 4.0     & 484      & Good, $\sim$1 arcsec \\
  \multicolumn{2}{l}{\bf SDSS\,J0106$-$0014:}\\
  2010-11-11  & ULTRACAM  & NTT      & $u'g'i'$ & 00:22:51 & 0.28--1.64 & 4.0     & 2494     & Good, $\sim$1 arcsec \\
  2010-11-23  & ULTRACAM  & NTT      & $u'g'i'$ & 00:30:57 & 0.48--0.60 & 5.0     & 1949     & Good, $\sim$1.5 arcsec \\
  2010-11-26  & ULTRACAM  & NTT      & $u'g'i'$ & 00:40:45 & 0.85--1.25 & 2.5     & 1159     & Excellent, $<$1 arcsec \\
  2010-11-26  & ULTRACAM  & NTT      & $u'g'i'$ & 04:22:13 & 0.66--1.20 & 4.0     & 996      & Excellent, $<$1 arcsec \\
  2010-11-27  & ULTRACAM  & NTT      & $u'g'i'$ & 00:29:50 & 0.52--1.13 & 2.5     & 1868     & Excellent, $<$1 arcsec \\
  2010-12-17  & ULTRACAM  & NTT      & $u'g'i'$ & 00:42:45 & 0.86--1.14 & 3.0     & 671      & Good, $\sim$1 arcsec \\
  2011-11-01  & ULTRACAM  & WHT      & $u'g'r'$ & 20:30:02 & 0.86--1.24 & 4.0     & 703      & Excellent, $<$1 arcsec \\
  2011-11-02  & ULTRACAM  & WHT      & $u'g'r'$ & 20:59:54 & 0.87--1.22 & 3.0     & 843      & Average, $\sim$1.5 arcsec \\
  2012-09-06  & ULTRACAM  & WHT      & $u'g'r'$ & 23:40:07 & 0.81--1.18 & 4.0     & 753      & Good, $\sim$1.5 arcsec \\
  2012-09-07  & ULTRACAM  & WHT      & $u'g'r'$ & 01:40:32 & 0.80--1.04 & 2.0     & 926      & Excellent, $<$1 arcsec \\
  2012-10-10  & ULTRACAM  & WHT      & $u'g'r'$ & 21:57:26 & 0.91--1.08 & 3.1     & 401      & Excellent, $<$1 arcsec \\
  2016-11-08  & ULTRACAM  & NTT      & $u'g'r'$ & 01:04:45 & 0.92--1.07 & 4.0     & 278      & Excellent, $<$1 arcsec \\
  \multicolumn{2}{l}{\bf SDSS\,J0110+1326:}\\
  2007-10-21  & ULTRACAM  & WHT      & $u'g'i'$ & 02:51:57 & 0.86--1.07 & 1.2     & 4805     & Good, $\sim$1.5 arcsec \\
  2010-11-23  & ULTRACAM  & NTT      & $u'g'i'$ & 02:57:19 & 0.45--0.56 & 3.0     & 1140     & Good, $\sim$1.5 arcsec \\
  2012-09-09  & ULTRACAM  & WHT      & $u'g'i'$ & 00:35:56 & 0.97--1.04 & 2.0     & 858      & Good, $\sim$1 arcsec \\
  2012-10-09  & ULTRACAM  & WHT      & $u'g'r'$ & 22:39:02 & 0.91--1.04 & 2.0     & 1829     & Good, $\sim$2 arcsec \\
  2015-09-19  & ULTRACAM  & WHT      & $u'g'r'$ & 04:33:21 & 0.91--1.02 & 1.8     & 1764     & Average, $\sim$1.5 arcsec \\
  2016-11-08  & ULTRACAM  & NTT      & $u'g'r'$ & 01:30:59 & 0.97--1.03 & 2.0     & 936      & Excellent, $<$1 arcsec \\
  \hline
  \multicolumn{9}{r}{\it continues on the next page...}\\
\end{tabular}
\end{table*}

\begin{table*}
 \centering
  \begin{tabular}{@{}lcccccccc@{}}
  \hline
  Date at     &Instrument &Telescope &Filter(s) &Start     &Orbital     &Exposure &Number of &Conditions              \\
  start of run&           &          &          &(UT)      &phase       &time (s) &exposures &(Transparency, seeing)  \\
  \hline
  \multicolumn{2}{l}{\bf SDSS\,J0314+0206:}\\
  2012-10-14  & ULTRACAM  & WHT      & $u'g'r'$ & 02:42:14 & 0.94--1.05 & 1.6     & 1712     & Good, $\sim$1.5 arcsec \\
  2016-11-09  & ULTRACAM  & NTT      & $u'g'r'$ & 05:00:17 & 0.93--1.07 & 3.5     & 1040     & Excellent, $<$1 arcsec \\
  \multicolumn{2}{l}{\bf SDSS\,J1021+1744:}\\
  2015-01-17  & ULTRACAM  & WHT      & $u'g'r'$ & 01:12:30 & 0.78--1.29 & 4.0     & 1563     & Good, $\sim$1 arcsec \\
  2015-05-19  & ULTRACAM  & WHT      & $u'g'z'$ & 21:13:23 & 0.89--1.09 & 9.9     & 245      & Good, $\sim$2 arcsec \\
  2015-05-20  & ULTRACAM  & WHT      & $u'g'i'$ & 21:11:18 & 0.01--0.48 & 5.0     & 1160     & Good, $\sim$1.5 arcsec \\
  \multicolumn{2}{l}{\bf SDSS\,J1028+0931:}\\
  2014-11-27  & ULTRASPEC & TNT      & $g'$     & 21:14:20 & 0.90--1.06 & 1.0     & 3163     & Good, $\sim$1.5 arcsec \\
  2014-12-01  & ULTRASPEC & TNT      & $g'$     & 20:48:21 & 0.85--1.04 & 1.0     & 3943     & Good, $\sim$2 arcsec \\
  2015-01-06  & ULTRASPEC & TNT      & $g'$     & 20:08:27 & 0.92--1.07 & 1.0     & 3056     & Good, $\sim$2 arcsec \\
  2015-02-25  & ULTRASPEC & TNT      & $g'$     & 15:33:21 & 0.86--1.07 & 1.0     & 4377     & Good, $\sim$2 arcsec \\
  2015-03-03  & ULTRASPEC & TNT      & $g'$     & 18:21:00 & 0.88--1.09 & 1.0     & 4199     & Good, $\sim$1.5 arcsec \\
  2016-03-17  & ULTRASPEC & TNT      & $g'$     & 19:02:34 & 0.85--1.05 & 1.0     & 4091     & Average, $\sim$1.5 arcsec \\
  \multicolumn{2}{l}{\bf SDSS\,J1123$-$1155:}\\
  2014-01-28  & ULTRASPEC & TNT      & $KG5$    & 17:19:39 & 0.92--1.04 & 1.5     & 5404     & Good, $\sim$1.5 arcsec \\
  \multicolumn{2}{l}{\bf SDSS\,J1307+2156:}\\
  2015-02-28  & ULTRASPEC & TNT      & $KG5$    & 21:18:08 & 0.91--1.05 & 4.8     & 542      & Good, $\sim$2 arcsec \\
  \multicolumn{2}{l}{\bf SDSS\,J1329+1230:}\\
  2010-04-21  & ULTRACAM  & NTT      & $u'g'i'$ & 05:18:12 & 0.76--1.50 & 3.0     & 1738     & Excellent, $<$1 arcsec \\
  2010-04-22  & ULTRACAM  & NTT      & $u'g'i'$ & 03:44:44 & 0.31--0.53 & 3.8     & 410      & Good, $\sim$2 arcsec \\
  2010-04-22  & ULTRACAM  & NTT      & $u'g'i'$ & 05:20:42 & 0.13--0.70 & 3.8     & 1041     & Good, $\sim$2 arcsec \\
  2010-04-23  & ULTRACAM  & NTT      & $u'g'i'$ & 01:06:14 & 0.30--1.12 & 3.8     & 1496     & Good, $\sim$1.5 arcsec \\
  2010-04-23  & ULTRACAM  & NTT      & $u'g'i'$ & 04:05:44 & 0.85--1.13 & 3.8     & 531      & Good, $\sim$1.5 arcsec \\
  2010-04-23  & ULTRACAM  & NTT      & $u'g'i'$ & 06:01:07 & 0.83--1.30 & 3.8     & 861      & Good, $\sim$1.5 arcsec \\
  2010-04-25  & ULTRACAM  & NTT      & $u'g'i'$ & 04:49:24 & 0.91--1.65 & 3.9     & 1315     & Average, $\sim$1.5 arcsec \\
  2010-04-28  & ULTRACAM  & NTT      & $u'g'r'$ & 02:48:35 & 0.93--1.15 & 3.9     & 404      & Good, $\sim$1.5 arcsec \\
  \multicolumn{2}{l}{\bf SDSS\,J2235+1428:}\\
  2010-11-12  & ULTRACAM  & NTT      & $u'g'i'$ & 00:14:36 & 0.31--1.08 & 5.0     & 1933     & Good, $\sim$1 arcsec \\
  2010-11-15  & ULTRACAM  & NTT      & $u'g'i'$ & 01:06:38 & 0.32--0.61 & 5.0     & 717      & Good, $\sim$2 arcsec \\
  2011-11-01  & ULTRACAM  & WHT      & $u'g'r'$ & 21:20:33 & 0.96--1.13 & 5.0     & 351      & Excellent, $<$1 arcsec \\
  2012-09-07  & ULTRACAM  & WHT      & $u'g'r'$ & 00:33:05 & 0.87--1.06 & 4.0     & 598      & Good, $\sim$1.5 arcsec \\
  2012-09-10  & ULTRACAM  & WHT      & $u'g'r'$ & 21:57:26 & 0.81--1.09 & 4.0     & 851      & Good, $\sim$2 arcsec \\
  2015-09-19  & ULTRACAM  & WHT      & $u'g'r'$ & 20:44:26 & 0.89--1.06 & 4.0     & 551      & Average, $\sim$2 arcsec \\
  2016-11-09  & ULTRACAM  & NTT      & $u'g'r'$ & 01:14:41 & 0.91--1.08 & 4.1     & 512      & Excellent, $<$1 arcsec \\
  \multicolumn{2}{l}{\bf WD\,1333+005:}\\
  2014-02-02  & ULTRASPEC & TNT      & $KG5$    & 18:54:41 & 0.84--1.13 & 1.5     & 1998     & Good, $\sim$1.5 arcsec \\
  2015-02-23  & ULTRASPEC & TNT      & $KG5$    & 18:28:12 & 0.71--1.10 & 3.0     & 1384     & Good, $\sim$1.5 arcsec \\
  2015-03-30  & ULTRASPEC & TNT      & $KG5$    & 15:59:01 & 0.86--1.09 & 3.5     & 688      & Good, $\sim$2 arcsec \\
  2016-03-19  & ULTRASPEC & TNT      & $KG5$    & 16:13:42 & 0.76--1.10 & 4.0     & 900      & Average, $\sim$2 arcsec \\
  \hline
\end{tabular}
\end{table*}

\section{Journal of spectroscopic observations}

\begin{table*}
 \centering
  \caption{Journal of X-shooter spectroscopic observations. We have not
    included the NIR exposure times or number of exposures for objects with
    very low signal-to-noise ratios in this arm, since this data was discarded.}
  \label{tab:speclog}
  \begin{tabular}{@{}lcccccccc@{}}
  \hline
  Date at     & Start    & Orbital    & Exposure time (s)   & Number of exposures & Conditions              \\
  start of run& (UT)     & phase      & UVB/VIS/NIR         & UVB/VIS/NIR         & (Transparency, seeing)  \\
  \hline
  \multicolumn{2}{l}{\bf CSS\,080502:}\\
  2014-03-04  & 01:18:17 & 0.07--0.39 & 368/410/300 & 10/9/13   & Good, $\sim$1 arcsec \\
  2014-03-05  & 00:30:13 & 0.54--0.98 & 368/418/300 & 14/12/18  & Good, $\sim$1.5 arcsec \\
  \multicolumn{2}{l}{\bf CSS\,09704:}\\
  2013-10-25  & 00:23:19 & 0.16--0.40 & 600/638/-   & 6/6/-     & Good, $\sim$1 arcsec \\
  2013-10-27  & 00:31:05 & 0.98--1.07 & 600/638/-   & 3/3/-     & Excellent, $<$1 arcsec \\
  2014-10-30  & 01:02:06 & 0.46--0.94 & 600/600/-   & 10/10/-   & Excellent, $<$1 arcsec \\
  \multicolumn{2}{l}{\bf CSS\,21357:}\\
  2014-04-21  & 03:12:16 & 0.30--1.02 & 600/301/200 & 23/42/71  & Excellent, $<$1 arcsec \\
  2014-04-22  & 02:28:15 & 0.20--0.32 & 600/301/200 & 5/9/15    & Excellent, $<$1 arcsec \\
  \multicolumn{2}{l}{\bf CSS\,40190:}\\
  2014-03-06  & 00:18:40 & 0.92--1.70 & 530/565/-   & 10/10/-   & Good, $\sim$1.5 arcsec \\
  2014-03-07  & 00:18:21 & 0.60--1.45 & 495/530/-   & 16/16/-   & Good, $\sim$1.5 arcsec \\
  \multicolumn{2}{l}{\bf SDSS\,J0024+1745:}\\
  2013-10-25  & 04:16:41 & 0.87--0.99 & 606/294/100 & 4/7/23    & Good, $\sim$1 arcsec \\
  2013-10-26  & 00:35:24 & 0.10--0.91 & 606/294/100 & 21/39/108 & Good, $\sim$1.5 arcsec \\
  2014-10-31  & 01:44:41 & 0.99--1.07 & 600/300/100 & 3/5/17    & Excellent, $<$1 arcsec \\
  \multicolumn{2}{l}{\bf SDSS\,J0106$-$0014:}\\
  2013-10-27  & 02:11:21 & 0.52--1.70 & 480/515/-   & 16/16/-   & Good, $\sim$1.5 arcsec \\
  \multicolumn{2}{l}{\bf SDSS\,J0110+1326:}\\
  2010-10-01  & 03:08:20 & 0.17--0.24 & 300/337/385 & 6/6/6     & Good, $\sim$1.5 arcsec \\
  2013-10-25  & 03:00:25 & 0.68--0.81 & 300/338/386 & 11/10/10  & Good, $\sim$1 arcsec \\
  2013-10-26  & 04:44:29 & 0.91--1.12 & 300/338/386 & 18/16/16  & Good, $\sim$1.5 arcsec \\
  2013-10-27  & 01:10:42 & 0.47--0.56 & 300/338/386 & 9/8/8     & Good, $\sim$1.5 arcsec \\
  \multicolumn{2}{l}{\bf SDSS\,J0314+0206:}\\
  2013-10-26  & 06:42:33 & 0.36--0.64 & 360/398/386 & 19/17/19  & Good, $\sim$1.5 arcsec \\
  2014-10-30  & 04:12:03 & 0.67--1.29 & 410/400/440 & 32/32/34  & Excellent, $<$1 arcsec \\
  \multicolumn{2}{l}{\bf SDSS\,J1021+1744:}\\
  2014-03-04  & 02:46:35 & 0.54--0.84 & 645/615/300 & 6/6/15    & Good, $\sim$1.5 arcsec \\
  2014-03-05  & 02:20:18 & 0.53--1.67 & 645/615/300 & 20/20/44  & Good, $\sim$1.5 arcsec \\
  \multicolumn{2}{l}{\bf SDSS\,J1028+0931:}\\
  2014-03-06  & 01:51:24 & 0.71--1.26 & 408/397/254 & 19/19/33  & Good, $\sim$1 arcsec \\
  2014-04-20  & 00:28:02 & 0.92--1.09 & 408/397/254 & 8/8/14    & Excellent, $<$1 arcsec \\
  2014-04-22  & 00:26:59 & 0.42--0.70 & 408/397/254 & 12/12/20  & Excellent, $<$1 arcsec \\
  \multicolumn{2}{l}{\bf SDSS\,J1123$-$1155:}\\
  2014-03-07  & 03:20:47 & 0.57--0.65 & 495/530/270 & 10/10/20  & Good, $\sim$1.5 arcsec \\
  2014-04-19  & 00:05:54 & 0.30--0.45 & 495/530/270 & 20/17/33  & Good, $\sim$1 arcsec \\
  2014-04-20  & 23:39:22 & 0.87--1.05 & 495/530/270 & 21/19/40  & Excellent, $<$1 arcsec \\
  2014-04-21  & 23:52:26 & 0.19--0.20 & 495/530/270 & 3/3/6     & Excellent, $<$1 arcsec \\
  \multicolumn{2}{l}{\bf SDSS\,J1307+2156:}\\
  2014-04-20  & 01:41:05 & 0.63--1.38 & 400/300/100 & 32/39/127 & Excellent, $<$1 arcsec \\
  \multicolumn{2}{l}{\bf SDSS\,J1329+1230:}\\
  2010-04-05  & 02:38:17 & 0.76--1.63 & 300/338/-   & 16/16/-   & Good, $\sim$1.5 arcsec \\
  \multicolumn{2}{l}{\bf SDSS\,J2235+1428:}\\
  2010-09-30  & 23:38:14 & 0.40--1.35 & 300/337/-   & 27/27/-   & Good, $\sim$1.5 arcsec \\
  \multicolumn{2}{l}{\bf WD\,1333+005:}\\
  2014-03-06  & 05:11:05 & 0.54--1.60 & 366/408/300 & 27/24/36  & Good, $\sim$1 arcsec \\
  \hline
\end{tabular}
\end{table*}

\bsp
\label{lastpage}
\end{document}